\documentclass{article}

\usepackage{arxiv}
\usepackage[utf8]{inputenc} 
\usepackage[T1]{fontenc}    
\usepackage{hyperref}       
\usepackage{url}            
\usepackage{booktabs}       
\usepackage{amsfonts}       
\usepackage{microtype}      
\usepackage{graphicx}
\usepackage{natbib}

\usepackage{amsmath}
\usepackage{rotating}
\usepackage{caption}
\usepackage{bm}
\usepackage{algorithm}%
\usepackage{algorithmicx}%
\usepackage{caption}
\usepackage{subcaption}

\title{Reliability estimation in dependent stress–strength model with Clayton copula and modified Weibull margins}

\author{ \href{https://orcid.org/0000-0001-6457-0967}{\includegraphics[scale=0.06]{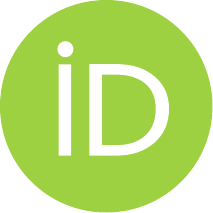}\hspace{1mm}Fatih K\i z\i laslan} \\
	Oslo, Norway\\
	\texttt{fkizilaslan@yahoo.com} \\

\renewcommand{\headeright}{}
\renewcommand{\undertitle}{}
\renewcommand{\shorttitle}{K\i z\i laslan F.}

\begin{document}
\maketitle

\begin{abstract}

Stress–strength models are widely used to assess the reliability of systems under uncertain conditions. While most studies assume independence between stress and strength variables, such an assumption may be unrealistic in many practical situations where these components are inherently dependent.
In this study, we investigate stress–strength reliability under a dependent framework, where both stress and strength variables follow modified Weibull distributions and their dependence is modeled via a Clayton copula. The proposed model allows distinct parameter sets, resulting in a flexible seven-parameter structure that extends Weibull-based models. We consider several estimation procedures for the model parameters and reliability, including two-step maximum likelihood, least squares, weighted least squares, and maximum product of spacings, with interval estimation obtained via asymptotic and bootstrap confidence intervals. The performance of the proposed estimators is evaluated through an extensive Monte Carlo simulation study under various parameter configurations and sample sizes.
Finally, the applicability of the proposed model is illustrated using monthly occupancy data from Istanbul’s two largest dams, with the Clayton copula capturing their dependence structure. 
This application demonstrates how stress–strength reliability can inform water management decisions and mitigate inter-regional operational risks.

\end{abstract}


\section{Introduction}

Assessing the ability of a system to withstand external loads is a fundamental problem in reliability engineering. A widely adopted probabilistic framework for addressing this problem is the stress–strength model, which evaluates system performance by comparing the random stress applied to a component with its inherent strength.
Let $X$ denote the random variable representing the strength of the system and $Y$ denote the applied stress. The reliability of the system is then defined as $R := P(X > Y)$, representing the probability that the available strength exceeds the imposed stress. Within this framework, system failure occurs whenever the applied stress surpasses the strength, that is, when $Y>X$. Consequently, the reliability measure $R$ provides a natural and interpretable metric for quantifying system robustness under uncertain operating conditions.
The stress–strength concept was first introduced by \cite{birnbaum1956use} and subsequently extended by \cite{birnbaum1958distribution}. Since then, the model has become a central topic in reliability analysis, attracting sustained interest in both theoretical and applied research. A comprehensive overview of its developments and applications can be found in the monograph by \cite{kotz2003stress}. Owing to its flexibility and interpretability, the stress–strength framework has been successfully applied in numerous areas, including quality control, structural engineering, seismology, medicine, and econometric modeling.

In the existing literature, stress–strength models are generally developed under the assumption that the stress and strength variables are independent. Consequently, most studies have focused on statistical inference for stress–strength models within this independence framework and under various distributional assumptions. Nevertheless, in many real-world applications, the stress and strength variables may exhibit some form of dependence.
Dependencies between components or system units have been observed across a wide range of fields, including engineering, quality control, economics, and education \cite{domma2013copula}. For example, \cite{domma2012stress} formulated a stress–strength problem in household budget management, treating household consumption as the stress variable $Y$ and income as the strength variable $X$. In this context, the reliability $R=P(X>Y)$ can be interpreted as a measure of household financial fragility.

Despite extensive research on stress–strength models over the past two decades, their interpretation and practical implementation of these models often lack clear explanations. Notable applications of dependent stress–strength models include the economic example discussed above, as well as the multicomponent framework introduced by \cite{KizilaslanNadar2018}, where dam or reservoir capacities are used to assess drought conditions in a given region.
This latter application can be naturally extended to a dependent stress–strength setting. Let $X$ and $Y$ denote the occupancy rates of two dams supplying the same city. Since both reservoirs serve a common demand, their levels are likely to exhibit positive dependence.
Accounting for the dependence between the two dams’ reserve capacities facilitates more realistic comparisons, improves the representation of resource management, and yields a more informative assessment of water system reliability.

Understanding cases in which stress and strength variables are dependent is essential for accurate reliability modeling. While the literature on independent stress–strength models is extensive, studies addressing dependent stress–strength structures remain limited. Early work on dependent models dates back to $2005$, where bivariate beta and exponential distributions were used to model the joint behavior of stress and strength variables in \cite{nadarajah2005reliability, nadarajah2006reliability}. 
Subsequent studies \cite{nadar2015estimation, KizilaslanNadar2018}  investigated pairs of dependent strength variables under a common independent stress variable, employing bivariate Weibull and Kumaraswamy distributions. A notable limitation of these bivariate distribution approaches is the requirement that the marginal distributions belong to the same family, which restricts modeling flexibility.
To overcome this limitation, copula-based approaches have been proposed, allowing the margins to follow arbitrary distributions without necessarily being of the same type. 

According to \cite{domma2012stress}, the first applications of copulas to dependent stress–strength models appeared around $2009$. In particular, \cite{domma2012stress} investigated the dependent stress-strength model reliability using a Frank copula to model dependence, with marginal distributions following the Dagum. Similarly, \cite{domma2013copula} considered dependent stress–strength models with Burr margins and dependence modeled through the Farlie-Gumbel-Morgenstern copula. 
These studies contribute important model formulations and demonstrate applications to real data; however, they remain limited in addressing statistical inference, concentrating on model development rather than the estimation and inference of $R$.
Nevertheless, they represent key starting points for the use of copula-based structures in stress–strength modeling.

In recent years, there has been a growing of research on statistical inference for the reliability of dependent stress–strength models. Several studies have addressed different system configurations, marginal distributions, and dependence structures.
\cite{bai2018reliability} investigated the estimation of stress–strength reliability for a system composed of series-connected strength components following a Weibull distribution, subjected to a common external stress following an exponential distribution, with dependence modeled via a Gumbel copula. 
\cite{zhu2022reliability} considered a multicomponent stress–strength model in which the stress and strength variables follow Kumaraswamy and unit-Gompertz distributions, respectively, with dependence constructed through a Clayton copula.
\cite{cai2023reliability} studied a multicomponent system where the strength variable consists of a pair of dependent Weibull components linked via a Clayton copula. 
\cite{shang2024reliability} discussed the estimation of reliability for a dependent stress–strength model under ranked set sampling, assuming that the stress and strength variables follow Kumaraswamy and Weibull distributions, respectively, with dependence characterized by a Clayton copula.
More recently, \cite{zhang2025bayesian} investigated both classical and Bayesian estimation for dependent stress–strength reliability in a series–parallel system, assuming that stress and strength follow distributions from the proportional reversed hazard rate family with dependence modeled by a Clayton copula. 
\cite{zhang2025copula} examined reliability estimation for multicomponent parallel–series systems, considering dependence between stress and strength via a Clayton copula, with Weibull and Burr XII distributions for the respective components.

In the reliability literature, numerous lifetime distributions have been proposed to extend classical models such as the Weibull distribution in order to achieve greater flexibility in modeling diverse failure behaviors. One such model is the modified Weibull distribution (MWD), introduced by \cite{lai2003modified}, which has been widely used in reliability and survival analysis.
A random variable $X$ is said to follow a modified Weibull distribution if its cumulative distribution function (cdf) $F(x)$ and probability density function (pdf) $f(x)$ are given by
\begin{eqnarray*}
    F(x) &=& 1- \exp \big( -a x^b e^{\lambda x} \big), \label{MWD_cdf} \\
    f(x) &=& a (b + \lambda x) x^{b-1} e^{\lambda x} \exp \big( -a x^b e^{\lambda x} \big) \label{MWD_pdf},
\end{eqnarray*}
where $x>0$, $a>0$ is the scale parameter, $b \ge 0$ is a shape parameter, and
$\lambda \ge 0$ is an acceleration or flexibility parameter that controls how quickly the hazard grows over time.
Larger $\lambda$ values correspond to faster failure accumulation and higher fragility. Restricting $\lambda \ge 0 $ ensures a non-decreasing hazard function, which is particularly suitable for reliability applications.
We denote this distribution by $MWD(a,b,\lambda)$, that is $X \sim MWD(a,b,\lambda)$.
The MWD includes several well-known distributions as special cases. In particular, when $\lambda=0$, it reduces to the two-parameter Weibull distribution with $F(x) = 1- \exp(-a x^b)$. 
When $b=0$, it reduces to a type I extreme-value (or log-gamma) distribution with $F(x) = 1- \exp(-a  e^{\lambda x} )$.
The stress–strength reliability problem under the MWD has received some attention in the literature. 
\cite{soliman2013reliability} investigated statistical inference for the stress–strength reliability when both stress and strength variables independently follow the MWD, sharing common shape and acceleration parameters while having distinct scale parameters.
Subsequently, \cite{goel2020estimation} extended this model under a progressive Type-II censoring scheme.

Motivated by the flexibility of the MWD, this study investigates the estimation of stress–strength reliability under a dependent framework. Specifically, we consider a model in which both stress and strength variables follow MWDs with distinct parameter sets, while their dependence structure is modeled using a Clayton copula.
Although several studies have examined dependent stress–strength models, the MWD has not yet been considered in this context, despite the fact that its special case--the two-parameter Weibull distribution--has been widely applied. 

The main contributions of this study are summarized as follows: (i) we propose a copula-based dependent stress–strength model with MWD margins that allows all parameters to be distinct, resulting in a flexible seven-parameter structure; (ii) we develop several estimation procedures for both the model parameters and the reliability $R$, including two-step maximum likelihood (ML), least squares (LS), weighted least squares (WLS), and maximum product of spacings (MPS) methods; (iii) we provide interval estimation of all parameters and $R$ using asymptotic confidence intervals based on ML, and bootstrap confidence intervals (BCIs) for all methods; (iv) we evaluate the performance of the proposed estimators through extensive Monte Carlo simulations and illustrate their practical applicability using a real data example; (v) the framework extends existing Weibull-based dependent stress–strength models by incorporating the additional flexibility of the MWD and relaxing the common-parameter assumption, which introduces additional computational challenges while better accommodating real-world systems; and (vi) the model reduces to the independent stress–strength case when the dependence parameter approaches zero, thereby generalizing the framework studied by \cite{soliman2013reliability}, in which stress and strength components independently follow MWD with common parameters.

The remainder of the paper is organized as follows. Section \ref{S:preliminaries} provides a brief overview of copula theory and introduces the proposed model, including its assumptions and the derivation of the stress-strength reliability $R$.
Section \ref{S:inference} presents the statistical inference procedures for the model, focusing on point and interval estimation of $R$.
In particular, the two-step ML, LS, WLS, and MPS methods are employed for point estimation, while asymptotic confidence intervals based on ML and bootstrap intervals for all methods are considered for interval estimation.
Section \ref{S:simulation} reports the results of a comprehensive Monte Carlo simulation study.
Section \ref{S:real_data} presents a real data application of the proposed methods based on observations from the two largest-capacity dams in Istanbul.
Finally, concluding remarks are presented in Section \ref{S:conclusion}.

\section{Preliminaries}{\label{S:preliminaries}}

\subsection{Copula}{\label{S:copula}}

For a $n$-dimensional continuous random vector $\mathbf{X}=(X_1,X_2,\dots, X_n)$ with joint cumulative distribution function (cdf) $H(x_1,x_2,\dots,x_n)$ and univariate marginal cdfs $F_1(x_1),F_2(x_2),\dots,F_n(x_n)$. 
Then, based on Sklar's Theorem there exist a unique $n$-dimensional copula function $C:[0,1]^n \rightarrow [0,1]$ satisfying 
\begin{equation*}
    H(\mathbf{x})=C \big( F_1(x_1), F_2(x_2),\dots,F_n(x_n) \big).
\end{equation*}
Let $c$ and $h$ be the corresponding joint probability density function (pdf) of $C$ and $H$, respectively, and $f_i$ be the corresponding pdf of $F_i, \; i=1,2,\dots,n$, we have 
\begin{eqnarray*}
    h(x_1,x_2,\dots,x_n) &=& \frac{\partial^nH(x_1,x_2,\dots,x_n)}
    {\partial{x_1} \partial{x_2} \dots \partial{x_n}} = 
    \frac{C \big( F_1(x_1), F_2(x_2),\dots,F_n(x_n) \big) }{\partial{F_1(x_1)} \partial{F_2(x_2)} \dots \partial{F_n(x_n)}} 
    \prod_{i=1}^{n}f_i(x_i) \notag  \\ 
    &=& c \big(F_1(x_1), F_2(x_2),\dots,F_n(x_n) \big) \prod_{i=1}^{n}f_i(x_i).
\end{eqnarray*}
For $2$-dimensional cases, let $h$, $f$, $g$ and $c$ be the pdf of the cdfs $H$, $F$, $G$ and $C$, respectively. Then, we have
\begin{eqnarray}
    h(x,y) = \frac{\partial H(x,y)}{\partial x \partial y} = \frac{\partial C(F(x), G(y))}{\partial x \partial y} = c \big( F(x), G(y) \big) f(x) g(y). \label{h_joint_pdf}
\end{eqnarray}
Several classes of copulas exist, including the Gaussian, t-, and Archimedean families. For a comprehensive introduction to copula theory and its applications, see \cite{nelsen2006introduction}. The Archimedean family, in particular, comprises many widely used copulas, including the independence (product) copula, Clayton, Gumbel, and Frank copulas, which are favored for their analytical tractability and flexibility in modeling diverse dependence structures.

In this study, we use the $2$-dimensional Clayton copula to depict the dependence of the stress and strength variables, which is given by
\begin{equation}
    C_{\theta}(u,v) = (u^{-\theta} + v^{-\theta} - 1)^{-1/\theta}, \; \theta >0 \label{ClaytonCopula}.
\end{equation}
Then, the joint pdf of $2$-dimensional Clayton copula is given as
\begin{equation}
    c_{\theta}(u,v) = (\theta +1)  u^{-(\theta + 1)} v^{-(\theta + 1)} \big( u^{-\theta} + v^{-\theta}-1 \big)^{-\left (\frac{1}{\theta} + 2 \right)}, \; \theta >0 \label{ClaytonCopula_pdf}.
\end{equation}
The first partial derivative of (\ref{ClaytonCopula}) with respect to $u$ can be written as 
\begin{equation}
    \frac{\partial C_{\theta}(u,v)}{\partial u} = u^{-(\theta + 1)}\big( u^{-\theta} + v^{-\theta}-1 \big)^{-\left (\frac{1}{\theta}+1 \right)}. \label{dC_u}
\end{equation}

Notice that the dependence between the associated random variables in the Clayton copula is governed by the parameter $\theta$, where $\theta \rightarrow 0$ corresponds to independence, and larger positive values of $\theta$ indicate stronger positive dependence, with $\theta \rightarrow \infty$ representing perfect dependence.
The strength of dependence can be summarized using Kendall’s $\tau$ coefficient, which measures the degree of concordance between two random variables by quantifying the difference between the probabilities of concordant and discordant pairs. For the Clayton copula, Kendall’s $\tau$ admits the closed-form expression as $\tau = \theta / (\theta + 2)$. 
This relationship establishes a one-to-one correspondence between the copula parameter $\theta$ and the overall strength of monotonic association, where $\tau =0$ represents independence, and values of $\tau$ closer to one indicate stronger positive dependence. 
Consequently, the Clayton copula parameter $\theta$ can be expressed in terms of Kendall’s $\tau$ as $\theta = 2\tau / (1-\tau)$ allowing the dependence parameter to be interpreted through an easily understandable rank-based measure and facilitating parameter calibration in empirical analyses. 
Additionally, the dependence parameter $\theta$ can be estimated by a semi-parametric technique method-of-moment based on the inversion of Kendall’s $\tau$. Then, the consistent estimator of $\theta$ is given by 
\begin{equation}
    \hat{\theta}_{\tau} = \frac{2 \hat{\tau}}{ 1-\hat{\tau} } ,\label{theta_Kendal_est}
\end{equation}
where $\hat{\tau}$ is the consistent estimator of the Kendall’s $\tau$ based on observed bivariate random sample.
This approach for estimating the copula dependence parameter $\theta$ has been used before in the similar studies such as \cite{bai2018reliability} and \cite{zhang2025bayesian}.

\subsection{Model assumptions}\label{S:model}
 
In this study, we assume that the strength $X$ and stress $Y$ variables are dependent which is constructed by using $2$-dimensional Clayton copula. Both components are distributed MWD without having any common parameters as $X \sim MWD(a_1,b_1,\lambda_1)$ and $Y \sim MWD(a_2,b_2,\lambda_2)$ with corresponding cdfs and pdfs $F_X(x;a_1,b_1,\lambda_1)$, \;$f_X(x;a_1,b_1,\lambda_1)$ and $G_Y(y;a_2,b_2,\lambda_2), \; g_Y(y;a_2,b_2,\lambda_2)$, respectively.
Based on the joint pdf $h(x,y)$ in (\ref{h_joint_pdf}), the stress-strength reliability $R$ is given as 
\begin{eqnarray*}
    R = P(X>Y) &=& \int_{0}^{\infty} \int_{0}^{x} h(x,y) \mathrm{d}y \mathrm{d}x 
    = \int_{0}^{\infty} \int_{0}^{x} \frac{\partial C^2(u,v)}{\partial u \partial v} \Biggr|_{ \begin{smallmatrix} u=F_X(x) \\ v = G_Y(y) \end{smallmatrix} } f_X(x) g_Y(y) \mathrm{d}y \mathrm{d}x \notag \\  
    &=& \int_{0}^{\infty}  \frac{\partial C(u,v)}{\partial u} \Biggr|_{ \begin{smallmatrix} u=F_X(x) \\ v = G_Y(x) \end{smallmatrix} } f_X(x) \mathrm{d}x.
\end{eqnarray*}
Under our model assumptions $X \sim MWD(a_1,b_1,\lambda_1)$, $Y \sim MWD(a_2,b_2,\lambda_2)$ with the $2$-dimensional Clayton copula from (\ref{ClaytonCopula}), $R$ becomes 
\begin{eqnarray}
    R &=& \int_{0}^{\infty}  F_X(x)^{-(\theta + 1)} \big(F_X(x)^{-\theta} + G_Y(x)^{-\theta}-1 \big) ^{-\left (\frac{1}{\theta}+1 \right)} f_X(x) \mathrm{d}x \notag \\
    &=& \int_{0}^{1} t^{-(\theta +1)} \big( t^{-\theta} + G_Y(F_{X}^{-1}(t))^{-\theta} -1 \big)^ {-\left(\frac{1}{\theta}+1 \right) } \mathrm{d}t, \label{reliability_clayton_MWD}
\end{eqnarray}
where $F_X(x) \equiv F_X(x;a_1,b_1,\lambda_1)  = 1- \exp(-a_1 x^{b_1} e^{\lambda_1 x})$ and $G_Y(y) \equiv G_Y(y;a_2,b_2,\lambda_2) = 1- \exp(-a_2 y^{b_2} e^{\lambda_2 y})$.
It is clear that $R$ has not an explicit form, and also it is not possible even when $X \equiv Y$ case. 
Notice that even for the independent stress and strength components case, i.e. $\theta \rightarrow 0$, there is no closed $R$ formula except $b_1=b_2, \; \lambda_1 = \lambda_2$ case as in \cite{soliman2013reliability}. Therefore, our model reliability formula as in (\ref{reliability_clayton_MWD}) also includes the independent components case under any assumptions of the distribution parameters.

In this study, we aim to examine the stress-strength reliability in a very general form, as expressed in (\ref{reliability_clayton_MWD}) without imposing any restrictions to the model parameters.
For this aim, a numerical integration method from \cite{Rprogram} statistical software is employed to the integral in (\ref{reliability_clayton_MWD}) for evaluation of $R$.
Notice that we have preferred to use the integral in the finite interval here to be able to have numerical stability for the integral calculations.
Additionally, we can reach an approximate value of $R$ using the Monte Carlo integration method based on a random generated sample as in \cite{curan2025statistical}. 
This approximation based on Monte Carlo integration method is given by
\begin{eqnarray}
    R &\approx&  \frac{1}{N} \sum_{i=1}^{N} T(x; \bm{\Omega_1},\bm{\Omega_2},\theta) = \tilde{R}, \label{R_approx}
\end{eqnarray}
where $\bm{\Omega_1}=(a_1,b_1,\lambda_1)$ and $\bm{\Omega_2}=(a_2,b_2,\lambda_2)$ represent the distributional parameters, $T(x; \bm{\Omega_1},\bm{\Omega_2}, \theta) =  F_X(x)^{-(\theta + 1)} \big( F_X(x)^{-\theta} + G_Y(x)^{-\theta}-1 \big)^{-\left (\frac{1}{\theta}+1 \right)}$, and $x$ is a random sample generated from $MWD(a_1,b_1,\lambda_1)$. 
This approach works sufficiently enough depends on the generated sample size $N$. 
It is known that $\Tilde{R} \rightarrow  R $ as $N \rightarrow \infty$ by the law of large numbers.
Along this study with all computations, we prefer to use the exact value of $R$ in (\ref{reliability_clayton_MWD}) using numerical integration method in \cite{Rprogram}. 
We have not generally encounter any numerical error for the evaluation of this integral, even if it arises, the suggested numerical integration method offers very good approximate value for around $N = 10^5$.

\section{Statistical inference of $R$}\label{S:inference}

In this section, we focus on the parameter and interval estimations of underlying distributions parameters of our model along with the stress-strength reliability $R$.  
Based on our model assumptions, suppose $n$ systems are put on a life-testing, and the observed dependent samples of the system strength, $X_i$, and stress variables,$Y_i$, are $(x_i,y_i)\; i=1,\dots, n$. 
Then, the likelihood function based on the observed samples is 
\begin{equation*}
 L(\bm{\Omega_1},\bm{\Omega_2},\theta; \mathbf{x},\mathbf{y}) = \prod_{i=1}^{n} c_{\theta} \big( F_{X}(x_i; \bm \Omega_1), G_Y(y_i; \bm \Omega_2) \big) f_X(x_i; \bm \Omega_1) g_Y(y_i; \bm \Omega_2).
\end{equation*}
The corresponding log-likelihood function for observations of strength and stress can be written as a summation of three terms $\ell_1, \; \ell_2$ and $\ell_3$ as 
\begin{equation}
    \ell(\bm{\Omega_1},\bm{\Omega_2},\theta; \mathbf{x},\mathbf{y}) =
    \ell_1(\bm{\Omega_1}; \mathbf{x}) + \ell_2(\bm{\Omega_2}; \mathbf{y}) + \ell_3(\bm{\Omega_1},\bm{\Omega_2},\theta; \mathbf{x},\mathbf{y}) \label{log-lik},
\end{equation}
where 
\begin{align}
    \ell_1(\bm{\Omega_1}; \mathbf{x}) &=  
    n \log(a_1) + \sum_{i=1}^{n} \log(b_1+\lambda_1 x_i)  + (b_1-1)  \sum_{i=1}^{n} \log x_i + \lambda_1  \sum_{i=1}^{n} x_i - a_1  \sum_{i=1}^{n}x_i^{b_1} e^{\lambda_1 x_i},  \notag \\ 
  \ell_2(\bm{\Omega_1}; \mathbf{y}) &= 
  n \log(a_2) + \sum_{i=1}^{n} \log(b_2+\lambda_2 y_i) + (b_2-1)  \sum_{i=1}^{n} \log y_i + \lambda_2  \sum_{i=1}^{n} y_i - a_2  \sum_{i=1}^{n} y_i^{b_2} e^{\lambda_2 y_i}, \notag\\ 
  \ell_3(\bm{\Omega_1},\bm{\Omega_2},\theta; \mathbf{x},\mathbf{y}) &=
  \sum_{i=1}^{n} \log \big( c_{\theta} \big( F_{X}(x_i; \bm \Omega_1), G_Y(y_i; \bm \Omega_2) \big) \big) \label{log-lik-3}.
\end{align}

\subsection{Maximum likelihood estimation of $R$}\label{Ss:MLE}

In the literature, there are different approaches for obtaining ML estimates of the unknown parameters in the similar studies. 
The first one is the classical MLE over unknown seven parameters simultaneously, based on  the full log-likelihood function given in (\ref{log-lik}). 
Since the Clayton copula density contains highly nonlinear terms and also strong interaction between marginal and dependence parameters, this way brings out solving a high-dimensional nonlinear optimization problem. It then may lead to numerical instability and convergence issues, particularly in small samples.
The second way as used in \cite{bai2018reliability}, firstly to obtain the estimate of copula parameter $\theta$ via Kendal's $\tau$ estimate, then employing maximization of the log-likelihood function using this $\hat{\theta}_{\tau}$ as fixed there.
 
The last approach approach is a two-step MLE also known as  the method of inference function for margins (IFM) which was fist introduced by \cite{xu1996statistical}.
In the first step, only the marginal parameters are estimated. In the second step, the copula parameters are estimated using the estimated marginal parameters obtained in the first step. The advantage over the full MLE approach, is that one does not have a single large problem, but multiple smaller optimization problems which are easier and faster to solve. 
Due to a similar performance but a more tractable and easier formulation, IFM is widely preferred over full MLE, for more details see \cite{joe1996estimation} and \cite{joe2005asymptotic}.
This two-step MLE method is also preferred in the similar reliability studies such as \cite{zhu2022reliability, shang2024reliability, zhang2025copula}.
Therefore, we use the two-step MLE method for obtaining the MLEs of our model. 
On the other hand, the existence and uniqueness of the MLEs of MWD parameters was presented by \cite{jiang2010mles}.
In our case, since the first step of the two-step MLE method includes the log-likelihood functions of MWD $l_1(\bm{\Omega_1}; \mathbf{x})$ and $l_2(\bm{\Omega_2}; \mathbf{y})$, it allows us to reduce optimization errors for MLE of the parameters in the first step.
Hence, this also becomes another important reason for choosing the two-step MLE method for our case.

In the first-step, we maximize the log-likelihood functions $l_1(\bm{\Omega_1}; \mathbf{x})$ and $l_2(\bm{\Omega_2},\mathbf{y})$, separately, in which their parameters are independent each other, for MLEs of $\bm{\Omega_1}=(a_1,b_1,\lambda_1)$ and  $\bm{\Omega_2}=(a_2,b_2,\lambda_2)$  as $\widehat{\bm{\Omega}}_1=(\hat{a}_1, \widehat{b}_1,\hat{\lambda}_1)$ and $\widehat{\bm{\Omega}}_2=(\hat{a}_2, \widehat{b}_2,\hat{\lambda}_2)$. Then, the MLEs of $a_1$ and $a_2$ have explicit forms as 
$$\hat{a}_1= \frac{n}{ \sum_{i=1}^n x_i^{b_1} e^{\lambda_1 x_i} }, \; \hat{a}_2= \frac{n}{ \sum_{i=1}^n y_i^{b_2} e^{\lambda_2 y_i} },
$$
and the remaining parameters are obtained simultaneously maximizing $\ell_1(\hat{a}_1, b_1, \lambda_1)$ and $\ell_2(\hat{a}_2, b_2, \lambda_2)$. 
In the second step, the MLE of the dependence parameter $\theta$, $\hat{\theta}$, is obtained by maximizing the log-likelihood function in (\ref{log-lik-3})  
\begin{align*}
 \ell_3(\theta; \widehat{\bm{\Omega}}_1, \widehat{\bm{\Omega}}_2, \mathbf{x},\mathbf{y}) & 
 = n \log(\theta + 1) - (\theta + 1) \sum_{i=1}^{n} \log(F_{X}(x_i; \widehat{\bm \Omega}_1) G_Y(y_i; \widehat{ \bm \Omega}_2) ) \notag \\
  &- \Big( \frac{1}{\theta} + 2\Big) \sum_{i=1}^{n} \log \big( (F_{X}(x_i; \widehat{ \bm \Omega}_1))^{-\theta} + (G_Y(y_i; \widehat{\bm \Omega}_2))^{-\theta} - 1 \big),
\end{align*}
with respect to the dependence parameter $\theta$.
In both steps, we maximize the non-linear equations using the limited-memory Broyden-Fletcher-Goldfarb-Shanno algorithm with box constraints (L-BFGS-B) in R statistical software \cite{Rprogram} via \texttt{optim} function. 
After obtaining the MLE of the parameters $(\hat{a}_1, \hat{b}_1, \hat{\lambda}_1)$, $(\hat{a}_2, \hat{b}_2, \hat{\lambda}_2)$ and $\hat{\theta}$, the MLE of $R$, denoted by $\hat{R}$, is obtained by substituting these estimates into (\ref{reliability_clayton_MWD}).

\subsection{Asymptotic confidence interval of $R$}\label{Ss:ACI}

Asymptotic confidence interval (ACI) for the model parameters and $R$ are constructed by using the asymptotic normality of MLE.  
When the copula dependence parameter has been considered as unknown parameter case, although ACI of the interested parameters has been presented in the similar studies such as \cite{zhu2022reliability, zhang2025copula}, there is no details about the implementation of it.
Due to used two-step MLE method along with the model complexity, constructing ACI of the parameters is not a straightforward approach especially for the copula parameter and any function depends on the model parameters.
That is why we follow the approach given in \cite{joe2005asymptotic} to obtain the ACIs of the model parameters $\bm{\Omega}_1, \bm{\Omega}_2, \theta$ and $R$.

As mentioned in \cite{joe2005asymptotic}, the ACI of $\bm{\Omega}_1$ and  $\bm{\Omega}_2$ are constructed by directly using the Fisher information matrix of corresponding univariate likelihood functions, namely $\ell_1(\bm{\Omega_1}; \mathbf{x})$ and $\ell_2(\bm{\Omega_2}; \mathbf{y})$. 
Due to our model complexity instead of calculating expectations for the Fisher information matrix, we prefer to use an observed information matrix which serves as a consistent estimator.
$\bm{\mathcal{I}}_1(\widehat{\bm{\Omega}}_1)= \big[ \mathcal{I}_{1,ij}\big] \big|_{\bm{\Omega}_1 = \widehat{\bm \Omega}_1} =\big[-\partial^2 \ell_1/\partial\Omega_{1i} \partial\Omega_{1j} \big] \big|_{\bm{\Omega}_1 = \widehat{\bm \Omega}_1}$, $i, j=1,2,3$ and 
$\bm{\mathcal{I}}_2(\widehat{\bm{\Omega}}_2)= \big[ \mathcal{I}_{2,ij}\big]\big|_{\bm{\Omega}_2 = \widehat{\bm \Omega}_2}  = \big[-\partial^2 \ell_2/\partial\Omega_{2i} \partial\Omega_{2j} \big] \big|_{\bm{\Omega}_2 = \widehat{\bm \Omega}_2} $, $i, j=1,2,3$ are the observed information matrices, namely negative of the Hessian matrices, corresponding to the log-likelihood functions $\ell_1$ and $\ell_2$, respectively.
For this aim, we employ the numerical Hessian matrices based on the optimized log-likelihood functions. Then, for $0 \leq \gamma \leq 1$, the $100(1-\gamma)\%$ ACIs of $\bm{\Omega_1}$ and $\bm{\Omega_2}$ are given as 
\begin{align*}
    \bigg( \widehat{\Omega}_{1i} - z_{\gamma /2} \sqrt{\mathcal{I}_{1,ii}^{-1}}, \;  \widehat{\Omega}_{1i} + z_{\gamma /2} \sqrt{\mathcal{I}_{1,ii}^{-1}} \bigg), \; i=1,2,3, \\
    \bigg( \widehat{\Omega}_{2i} - z_{\gamma /2} \sqrt{\mathcal{I}_{2,ii}^{-1}}, \;  \widehat{\Omega}_{2i} + z_{\gamma /2} \sqrt{\mathcal{I}_{2,ii}^{-1}} \bigg), \; i=1,2,3,
\end{align*}
where $\bm \Omega_{1}=(\Omega_{11}, \Omega_{12}, \Omega_{13}) = (a_1,b_1, \lambda_1)$, $\bm \Omega_{2}=(\Omega_{21}, \Omega_{22}, \Omega_{23}) = (a_2,b_2,\lambda_2)$ and $\mathcal{I}_{i,..}^{-1}, i=1,2$ show the elements of the inverse matrix of $\bm{\mathcal{I}}_{i}(\widehat{\bm \Omega}_i), i=1,2$.

For the ACI of the copula dependence parameter $\theta$, 
$\hat{\theta}$ is also asymptotically normally distributed under the two-step MLE framework.  
By following the study of \cite{joe2005asymptotic}, the $100(1-\gamma)\%$ ACI of $\theta$ is given by  $\big( \hat{\theta} - z_{\gamma /2} \sqrt{\widehat{V}_{77}}, \; \hat{\theta} + z_{\gamma /2} \sqrt{\widehat{V}_{77}} \big)$ where
$\widehat{V}_{77}$ represents the asymptotic variance of $\theta$, given by the diagonal element of the covariance matrix $\mathbf{V}$ associated with $\theta$, evaluated at the maximum likelihood estimator $\widehat{\bm{\Theta}}$,
$\mathbf{V}= (-\mathbf{D}^{-1}) \mathbf{M} (-\mathbf{D}^{-1})^{T}$ represents the sandwich matrix (Godambe Information matrix) by \cite{godambe1960optimum}
\begin{equation*}
    -\mathbf{D} = 
    \begin{pmatrix}
       \bm{\mathcal{I}}_{1} & \mathbf{0} & 0\\
        \mathbf{0} &  \bm{\mathcal{I}}_{2} & 0 \\
        \mathcal{I}_{3,71} &\cdots \mathcal{I}_{3,76}& \mathcal{I}_{3,77}
    \end{pmatrix}, \;
    \mathbf{M} = 
    \begin{pmatrix}
        \bm{\mathcal{I}}_{1} & \mathbf{0} & 0\\
        \mathbf{0} &  \bm{\mathcal{I}}_{2} & 0 \\
        \mathbf{0} & \mathbf{0} & \mathcal{I}_{3,77}
    \end{pmatrix},
\end{equation*}
and $\mathcal{I}_{3,7j}= -\partial^2 \ell_3 / \partial \theta \partial \Theta_j$,  $\bm{\Theta}=(a_1,b_1,\lambda_1,a_2,b_2,\lambda_2, \theta)$ .
Notice that the diagonal elements of $\mathbf{V}$ related to parameters $\bm{\Omega}_1$ and $\bm{\Omega}_2$ are the same with $\mathcal{I}_{1,ii}^{-1}$ and $\mathcal{I}_{2,ii}^{-1}, \; i=1,2,3$. 

As the last step, the ACI of the reliability $R=g(\bm \Theta)$ can be obtained by using delta method. 
Given the asymptotic normality of the two-step ML estimator, $\widehat{\bm \Theta} \sim N(\bm \Theta, \mathbf{V})$, the variance of $\hat{R}$ is approximated by: $Var (\hat{R}) \approx \nabla g(\widehat{\bm{\Theta}})^T \;\mathbf{V}  \; \nabla g(\widehat{\bm{\Theta}}) $, where $\nabla g(\widehat{\bm{\Theta}}) $is the numerical gradient vector of the reliability $R$ with respect to $\bm{\Theta}$.
Notice that due to complexity of the reliability $R$ formula given by integral as in (\ref{reliability_clayton_MWD}), which has not a closed form and depends on seven model parameters, the numerical derivatives are used for computation of the ACI of $R$.
Because of that the obtained ACI could be considered as a kind of an approximate confidence interval besides that might be a reason of the lack of robustness confidence interval in some cases.
We therefore also present a parametric bootstrap approach for all the parameters considering used methods in Subsection \ref{Ss:bootstrapCI}, which might provide more robust confidence intervals.

\subsection{Least square and weighted least square estimations of $R$}\label{Ss:LSE}

We obtain the regression-based estimators LSE and WLSE for the unknown parameters of our model. It was originally suggested by \cite{swain1988least}. Since then it has been used in literature for different distributions under diverse assumptions.
The LS estimator is obtained by minimizing the sum of squared differences between the theoretical cdf and the empirical cdf evaluated at the ordered sample points, with respect to the parameter of interest.

We apply the same two-step procedure as in the MLE case for both the LSE and WLSE methods.
Suppose $X_{(1)}<X_{(2)}< \dots < X_{(n)}$ and $Y_{(1)}<Y_{(2)}< \dots < Y_{(n)}$ are the ordered samples of the random sample of strength and stress variables $X_1,\dots,X_n$ and $Y_1,\dots, Y_n$, respectively.
Then, based on the $x_{(1)}<x_{(2)}< \dots < x_{(n)}$ and $y_{(1)}<y_{(2)}< \dots < y_{(n)}$ ordered observed samples, LS estimates of our model parameters $\bm{\Omega_1}$ and $\bm{\Omega_2}$ are obtained minimizing following sum of squares:
\begin{equation}
    Q_1(\bm{\Omega_1}) = \sum_{i=1}^n \bigg( F(x_{(i)};a_1,b_1,\lambda_1) - \hat{F}_n(x_{(i)})  \bigg)^2 ,\; 
    Q_2(\bm{\Omega_2}) = \sum_{i=1}^n \bigg( G(y_{(i)};a_2,b_2,\lambda_2) - \hat{G}_n(y_{(i)})  \bigg)^2 
    \label{LS_eq}
\end{equation}
with respect to the parameter of interest. Note that we prefer to use the Benard’s approximation for the empirical cdf at the ordered observations
as $\hat{F}(x_{(i)}) = \hat{G}(y_{(i)}) = (i-0.3)/(n+0.4), \; i=1,\dots,n$.

In the second step, the LSE of Clayton copula parameter $\theta$ is obtained using the obtained LSEs $\bm{\widehat{\Omega}_{1,LSE}}=(\hat{a}_{1,LSE}, \hat{b}_{1,LSE}, \hat{\lambda}_{1,LSE})$ and $\bm{\widehat{\Omega}_{2,LSE}}=(\hat{a}_{2,LSE}, \hat{b}_{2,LSE},\hat{\lambda}_{2,LSE})$. 
For this aim, similar to the univariate case above, we use the bivariate cdf via copula function $C_{\theta}$ and the corresponding bivariate empirical cdf $\hat{H}_n$. Then, the LSE of $\theta$, denoted by $\hat{\theta}_{LSE} $, is obtained by minimizing the following equation 
\begin{equation}
      Q_3(\theta) = \sum_{i=1}^n \bigg( C_{\theta}(u_i,v_i;\theta) - \hat{H}_n(x_i,y_i) \bigg)^2 \label{LS_eq_theta}
\end{equation}
with respect to $\theta$, where $u_i= F(x_i; \hat{a}_{1,LSE}, \hat{b}_{1,LSE}, \hat{\lambda}_{1,LSE})$, $v_i= G(y_i; \hat{a}_{2,LSE}, \hat{b}_{2,LSE}, \hat{\lambda}_{2,LSE})$ and $\hat{H}_n(x_i,y_i) = \frac{1}{n}\sum_{j=1}^n \mathbf{I}(X_j \leq x_i, Y_j \leq y_i)$.
Then, the LSE of $R$, denoted by $\hat{R}_{LSE}$, is obtained by substituting the corresponding LSEs into (\ref{reliability_clayton_MWD}).

In the weighted version of the LSE method, each observation is assigned a weight based on the variance of the $i$-th order statistic from a sample of size $n$ drawn from a uniform distribution $U(0,1)$. Specifically, the weights are taken as the inverse of the variance of $U_{(i)}$, given by
\begin{equation}
    w_i = \frac{1}{Var(U_{(i)})} = \frac{(n+1)^2(n+2)}{i(n-i+1)}, \; i=1,\dots,n. \label{WLS_weight}
\end{equation}
Therefore, the sum of squares of WLSE method become as 
\begin{equation}
    Q_1^W(\bm{\Omega_1}) = \sum_{i=1}^n w_i \bigg( F(x_{(i)};a_1,b_1,\lambda_1) - \frac{i-0.3}{n+0.4}  \bigg)^2 , \;
    Q_2^W(\bm{\Omega_2}) = \sum_{i=1}^n w_i \bigg( G(y_{(i)};a_2,b_2,\lambda_2) - \frac{i-0.3}{n+0.4}  \bigg)^2,
    \label{WLS_eq}
\end{equation}
\begin{equation}
    Q_3^W(\theta) = \sum_{i=1}^n w_i \bigg( C_{\theta}(u_i,v_i;\theta) - \hat{H}_n(x_i,y_i) \bigg)^2. \label{WLS_eq_theta}
\end{equation}
As in the LSE case, firstly we obtain WLS estimates $\bm{\widehat{\Omega}_{1,WLSE}}= (\hat{a}_{1,WLSE}, \hat{b}_{1,WLSE}, \hat{\lambda}_{1,WLSE})$ and $\bm{\widehat{\Omega}_{2,WLSE}} = (\hat{a}_{2,WLSE}, \hat{b}_{2,WLS},\hat{\lambda}_{2,WLSE})$ minimizing $Q_1^W(\bm{\Omega_1})$ and $Q_2^W(\bm{\Omega_2})$ functions.
In the next, $\hat{\theta}_{WLSE}$ is obtained by minimizing $Q_3^W(\theta)$ using the obtained WLSEs in the first step.
Hence, the WLSE of $R$, denoted by $\hat{R}_{WLSE}$, is computed by substituting the corresponding WLSEs into (\ref{reliability_clayton_MWD}).

On the other hand, estimating the copula parameter $\theta$ using classical LSE or WLSE especially in small samples may produce unstable or extremely large estimates, because the objective function is highly nonlinear and depends on empirical bivariate CDFs. Due to this reason, we apply a second-order Taylor expansion for the copula function, and then this linearization allows us to have a closed-form estimate of $\theta$ from the equations of $Q_3(\theta)$ in (\ref{LS_eq_theta}) and $Q_3^W(\theta)$ in (\ref{WLS_eq_theta}), which is called one-step solution. 
Based on this approximation, the one-step LS and WLS estimates of $\theta$ are given by
\begin{align}
    \theta_{LSE,one-step} = \hat{\theta}_{\tau} - \frac{B}{2C}, \;
    \theta_{WLSE,one-step} = \hat{\theta}_{\tau} - \frac{B^*}{2C^{*}}. \label{theta_one_step_est} 
\end{align}
All details of the one-step estimates of $\theta$ along with the coefficients $B,\;C, \; B^*$ and $C^*$ are given in Appendix \ref{A:approx_LS_WLS}.

\subsection{Maximum product spacing estimation of $R$}\label{Ss:MPS}

The maximum product of spacings (MPS) method, proposed by \cite{cheng1983estimating}, constitutes an alternative estimation technique to MLE, particularly useful when the likelihood function is difficult to maximize directly.
The uniform spacings for our model strength and stress variables $F(x;a_1,b_1,\lambda_1)$ and  $G(y;a_2,b_2,\lambda_2)$ based on the ordered samples $(x_{(i)},y_{(i)}), \; i=1,\dots, n$ are defined as 
\begin{equation*}
    D_{1i}(a_1,b_1,\lambda_1) = F(x_{(i)};a_1,b_1,\lambda_1) - 
F(x_{(i-1)};a_1,b_1,\lambda_1),
\end{equation*}
and 
\begin{equation*}
    D_{2i}(a_2,b_2,\lambda_2) = G(y_{(i)};a_2,b_2,\lambda_2) - 
G(y_{(i-1)};a_2,b_2,\lambda_2), 
\end{equation*}
where $i=1,\dots, n+1$, $F(x_{(0)};a_1,b_1,\lambda_1) = G(y_{(0)};a_2,b_2,\lambda_2) =0$ and $F(x_{(n+1)};a_1,b_1,\lambda_1) = G(y_{(n+1)};a_2,b_2,\lambda_2) =1$.
Then, the MPS estimates of $(a_1,b_1,\lambda_1)$ and $(a_2,b_2,\lambda_2)$, denoted by $\widehat{\bm \Omega}_{1,MPS} = (\hat{a}_{1,MPS},\hat{b}_{1,MPS},\hat{\lambda}_{1,MPS})$ and $\widehat{\bm \Omega}_{2,MPS} = (\hat{a}_{2,MPS},\hat{b}_{2,MPS},\hat{\lambda}_{2,MPS})$, are obtained by maximizing the following equations
\begin{equation*}
    M_1(a_1,b_1,\lambda_1) = \sum_{i=1}^{n+1} \frac{1}{n+1} \log \big(D_{1i} (a_1,b_1,\lambda_1) \big), \; \; \; 
    M_2(a_2,b_2,\lambda_2) = \sum_{i=1}^{n+1} \frac{1}{n+1} \log \big(D_{2i} (a_2,b_2,\lambda_2) \big),
\end{equation*}
with respect to the related parameters, respectively. This is the first step of MPS estimates, then in the second step we obtain the MPS estimate of $\theta$, $\hat{\theta}_{MPS}$, maximizing the following equation using the copula function based on the obtained MPS estimates 
\begin{equation*}
    M_3(\theta) = \sum_{i=1}^{n+1} \frac{1}{n+1} \log \big(D_{3i} (\theta ) \big) 
\end{equation*}
where $ D_{3i} (\theta) = C_{\theta}\big( F(x_{(i)}; \widehat{\bm \Omega}_{1,MPS}), G(y_{(i)};\widehat{\bm \Omega}_{2,MPS} ); \theta \big) - 
    C_{\theta}\big( F(x_{(i-1)}; \widehat{\bm \Omega}_{1,MPS}), G(y_{(i-1)}; \widehat{\bm \Omega}_{2,MPS} ); \theta \big) $.

\subsection{Bootstrap confidence intervals of $R$}\label{Ss:bootstrapCI}

We construct the ACI of the parameters based on MLEs, however it is not possible to derive the  confidence interval directly for other methods LS, WLS and MPS estimates. That is why the parametric bootstrap percentile method is employed to be able to construct bootstrap confidence intervals (BCIs) for unknown parameters and $R$ as well based on these methods. 
Even the ACI can be constructed as mentioned in Section \ref{Ss:ACI}, in certain cases due to lack of enough observations it can be needed to have an alternative method. 
Additionally, as mentioned above ACI includes numerical gradient computations because of model complexity, then it could be raised robustness issues in the confidence interval for some cases.
Therefore, because of these reasons, we also obtain and present the BCI for the MLE method. 
Algorithm \ref{alg:bootstrap} is employed for obtaining BCIs of the interested parameters based on MLE, LSE, WLSE and MPS methods.

\begin{algorithm}
\caption{Bootstrap confidence interval algorithm}
\label{alg:bootstrap}
\begin{algorithmic}[1]
\State Compute the estimates of parameters as $\hat{a}_1, \hat{b}_1, \hat{\lambda}_1, \hat{a}_2, \hat{b}_2$ and  $\hat{\lambda}_2$ using one of the estimation methods: ML, LS, WLS, or MPS, based on the original dependent sample.
\State Generate a dependent bootstrap sample using the estimated parameters
$\hat{a}_1, \hat{b}_1, \hat{\lambda}_1, \hat{a}_2, \hat{b}_2$ and  $\hat{\lambda}_2$.
\State Compute the bootstrap estimates $\hat{a}_1^{*}, \hat{b}_1^{*}, \hat{\lambda}_1^{*}, \hat{a}_2^{*}, \hat{b}_2^{*}$, and  $\hat{\lambda}_2^{*}$ based on this bootstrap sample.
\State Using these bootstrap estimates, compute the bootstrap estimate of the reliability parameter $R$, denoted by $\hat{R}^{*}$.
\State Steps $2-4$ independently $B$ times to obtain $B$ bootstrap estimates of the parameters such as $\{ \hat{R}_1^{*}, \hat{R}_2^{*}, \ldots, \hat{R}_B^{*} \}$.
\State The two-sided $100(1-\gamma)\%$ BCI for $R$ is given by $ \big( \hat{R}^{*[\gamma/2]}, \hat{R}^{*[1-\gamma/2]} \big) $ where $\hat{R}^{*[p]}$ denotes the $p$th empirical quantile of the bootstrap estimates $\{\hat{R}_1^{*},\ldots,\hat{R}_B^{*}\}$. 
BCIs for the parameters $a_1, b_1, \lambda_1, a_2, b_2,$ and $\lambda_2$ are obtained in the same manner.

\end{algorithmic}
\end{algorithm}

\section{Simulation study}\label{S:simulation}

In this section, the performance of the proposed estimation methods is evaluated in terms of bias and mean squared error (MSE) for point estimation, and confidence interval length and coverage probability (CP) for interval estimation. Dependent random samples are generated according to Algorithm \ref{alg:data_gen}.
All simulation scenarios were implemented in the R statistical software (version 4.5.2; \cite{Rprogram}), and the R code for implementing the proposed model algorithms is available on GitHub at \url{https://github.com/fatihki/SSReliabilityClaytonMWD}.
\begin{algorithm}
\caption{Dependent stress-strength samples generation}
\label{alg:data_gen}
\begin{algorithmic}[1]
\State Generate $n$ independent uniform  vectors $\mathbf{u},\; \mathbf{w} \sim \text{Uniform}(0,1)$.
\State Compute $\mathbf{v} = c_u^{-1}(\mathbf{w})$, where $c_u = \partial C_{\theta}/ \partial u$ in (\ref{dC_u}) and  $c_u^{-1}(.)$ is the pseudo-inverse of $c_u(.)$.
\State Generate samples using the quantile functions as $\mathbf{x} = F_X^{-1}(\mathbf{u})$ and $\mathbf{y} = G_Y^{-1}(\mathbf{v})$.
\State Return the dependent stress–strength samples $(\mathbf{x}, \mathbf{y})$.
\end{algorithmic}
\end{algorithm}

We conduct the simulation study under two main settings. In the first setting, the true parameter values are taken as  $\bm{\Omega}_1=(a_1, b_1,\lambda_1)=(0.75,1.25,0.6)$, $\bm{\Omega}_2=(a_2, b_2,\lambda_2)=(2,1.5,0.25)$ along with the different copula parameters $\theta \in \{2,3,4,6 \}$ and sample size $n \in \{25,50,100 \}$. 
The corresponding true values of the stress-strength reliability $R$ are $0.76403, 0.81533, 0.85699$ and $0.91303$ for $\theta = 2,3,4$ and $6$, respectively.
For this setting, all results are based on $1000$ Monte Carlo replications.
The point estimation methods considered include ML, LS, WLS, and MPS for all model parameters and the reliability $R$.
For the copula parameter $\theta$ and the reliability $R$, the one-step variants of LS and WLS are specifically employed.
For interval estimation, $95 \%$ ACIs based on ML estimates and BCIs based on all methods are constructed. The BCIs are obtained using $B=1000$ bootstrap samples for each of the $1000$ simulated datasets. 
The point estimation results, together with the associated bias and MSE values for all parameters under the MLE, one-step LSE, one-step WLSE, and MPS methods, are presented in Tables \ref{table:estimates1}–\ref{table:estimates2} for $\theta = 2, 3$ and $\theta = 4, 6$, respectively.
The corresponding $95\%$ ACIs and BCIs, together with their interval lengths and coverage probabilities (CPs), are presented in Tables \ref{table:interval1}–\ref{table:interval2}.

\begin{figure}[t]
\centering
\includegraphics[scale=0.7]{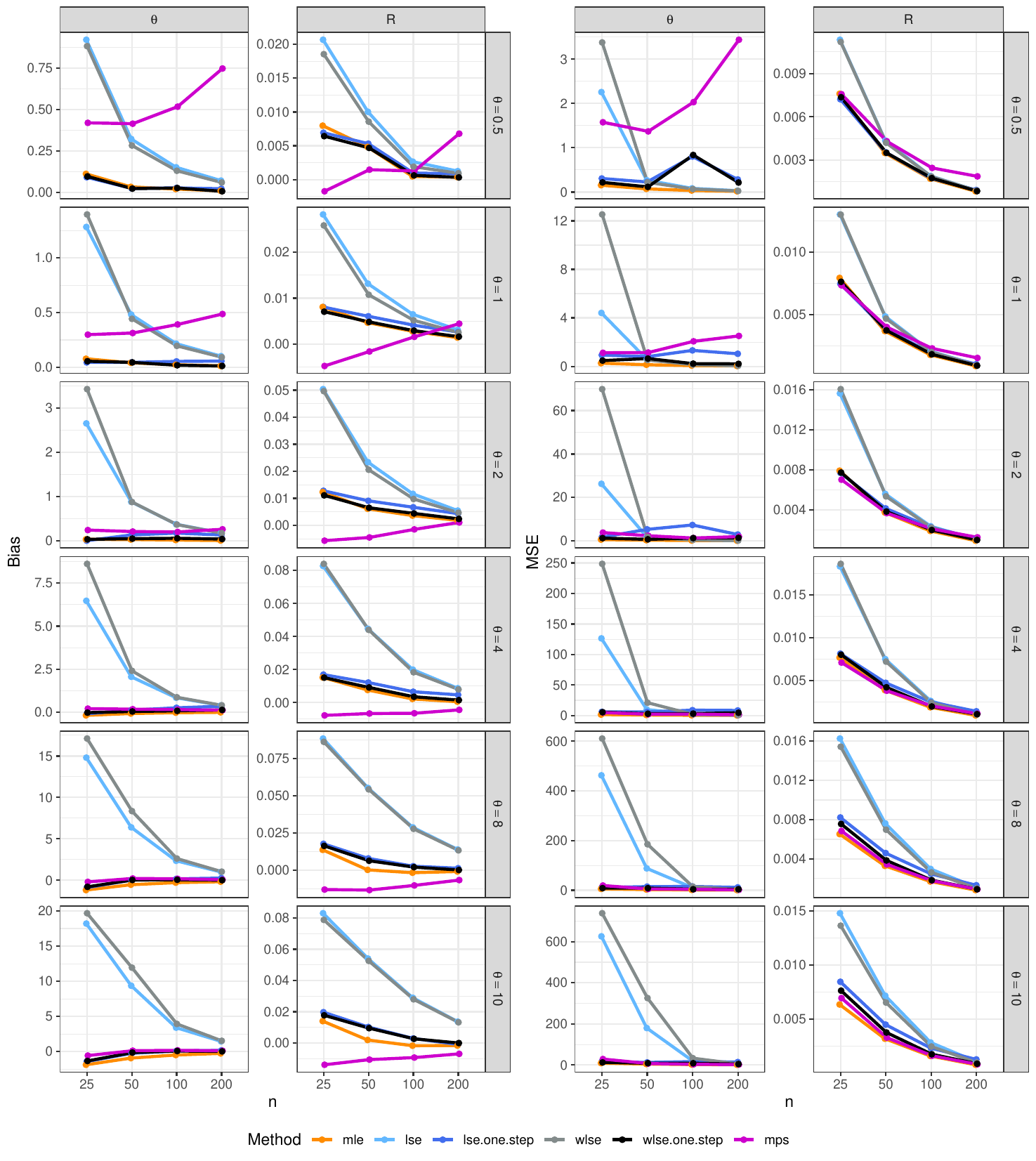}
\caption{Bias and MSE of the copula parameter $\theta$ and the stress-strength reliability $R$.} 
\label{bias_mse_Rtheta}
\end{figure}

In the second setting, we consider a broader range of values for the copula parameter $\theta$ and $n$. The true parameter values are taken as $\bm{\Omega}_1=(a_1, b_1,\lambda_1)=(0.6,0.75,1.4)$, $\bm{\Omega}_2=(a_2, b_2,\lambda_2)=(2.5,1.5,0.5)$ along with the different copula parameters $\theta \in \{0.5,1,2,4,8,10 \}$ and sample size $n \in \{25,50,100,200 \}$. The corresponding true values of the stress-strength reliability $R$ are $0.59389,0.59877, 0.61184, 0.64763, 0.71344 $ and $0.73588$ for $\theta = 0.5,1,2,4,8$ and $10$, respectively.
In this setting, we focus on point estimation of the model parameters and the reliability $R$ using the same methods in the first setting, along with the classical LS and WLS estimates for $\theta$ and $R$. 
The performance of the estimators is summarized through bias and MSE plots across different values of $\theta$ and $n$, facilitating a clear comparison of the methods.
All results are based on $5000$ Monte Carlo replications.
The performance of the copula parameter $\theta$ and the stress–strength reliability $R$, in terms of bias and MSE, under the MLE, LSE, one-step LSE, WLSE, one-step WLSE, and MPS methods, is illustrated in Figure \ref{bias_mse_Rtheta}. 
This figure facilitates a clear comparison between the classical and one-step versions of the LSE and WLSE methods.
For a clearer comparison of the other estimation methods after excluding the classical LSE and WLSE approaches, the corresponding figure is provided in Appendix Figure \ref{bias_mse_Rtheta_r1}.
The bias and MSE plots for the model parameters $\bm{\Omega}_1$ and $\bm{\Omega}_2$ are also presented in Appendix Figures \ref{bias_plot_parameters} and \ref{mse_plot_parameters}.

The distinction between the classical and one-step variants of the LSE and WLSE methods lies in the estimation of the copula parameter $\theta$ and the reliability $R$. In the one-step approaches, after obtaining the LSE (or WLSE) of $\bm{\Omega}_1$ and $\bm{\Omega}_2$ using (\ref{LS_eq}) (or \ref{WLS_eq} ), the copula parameter $\theta$ is estimated by using (\ref{theta_one_step_est}), and the reliability $R$ is subsequently computed by substituting these estimates into (\ref{reliability_clayton_MWD}). 

To solve the optimization problems arising in the MLE, LSE, WLSE, and MPS estimation methods, we employ the L-BFGS-B algorithm implemented in the R statistical software \cite{Rprogram} via the \texttt{optim} function.
For all simulations, the parameter space is restricted to $(0,\infty)$, and the optimization is initialized using randomly generated starting values from the interval $(0,1)$. For the copula parameter $\theta$, the initial value is taken as the estimate based on Kendall’s $\tau$, $\hat{\theta}_{\tau}$.

From Tables \ref{table:estimates1}–\ref{table:estimates2}, it is observed that as the sample size increases, the MSEs of all estimators decrease, while the biases also generally decrease, with a few exceptions in certain cases. The MSEs of the MLE and MPS methods for the parameters of $\bm{\Omega}_1$ and $\bm{\Omega}_2$ are very close to each other and are consistently smaller than those obtained from the LSE and WLSE methods. Furthermore, the WLSE/one-step WLSE method yields smaller MSE values than the LSE/one-step LSE method, indicating that LSE/one-step LSE generally produces the largest MSEs across all parameters. These patterns are also evident in the second simulation setting, as illustrated in Figure \ref{mse_plot_parameters}.
In addition, the MLE demonstrates the best overall performance in terms of both MSE and bias for the copula parameter $\theta$ and the reliability measure $R$.

From Tables \ref{table:interval1}–\ref{table:interval2}, it is observed that, as expected, the lengths of all confidence intervals decrease with increasing sample size.
As noted in Subsection \ref{Ss:ACI}, due to the complexity involved in deriving the ACIs based on the MLE, the CPs--particularly for the parameters of $\bm{\Omega}_1$ and $\bm{\Omega}_2$--may deviate substantially from the nominal level of $0.95$ in certain cases.
While the CP values for the copula parameter $\theta$ are generally close to the nominal level, those for the reliability measure $R$ tend to be slightly higher.
Therefore, to obtain more reliable and robust interval estimates, the use of bootstrap-based confidence intervals is recommended.

The bootstrap confidence intervals for the marginal strength and stress distribution parameters, $\bm{\Omega}_1$ and $\bm{\Omega}_2$, tend to be conservative, yielding coverage probabilities exceeding the nominal level (approximately $0.99$). This behavior can be attributed to the strong identifiability of the marginal parameters and the conservative nature of percentile-based bootstrap intervals that lead to inflated interval widths. In contrast, the copula parameter $\theta$ and the stress–strength reliability $R$, which are more challenging to estimate and involve nonlinear dependence structures, exhibit coverage probabilities closer to the nominal $95\%$ level.
Consequently, the CPs associated with the parameters of $\bm{\Omega}_1$ and $\bm{\Omega}_2$ are consistently higher than the nominal value of $0.95$ in Tables \ref{table:interval1}–\ref{table:interval2}.
Overall, the BCIs based on the MLE method demonstrate the best performance in terms of both interval length and CP for $\theta$ and $R$. For the marginal distribution parameters $\bm{\Omega}_1$ and $\bm{\Omega}_2$, however, the BCIs based on the MPS method provide comparatively better performance.

From Figure \ref{bias_mse_Rtheta}, it is observed that the LSE and WLSE methods yield the largest bias and MSE values for both the copula parameter $\theta$ and the reliability $R$ across different values of $\theta$ and sample sizes $n$, with the exception of the case $\theta = 0.5$.
These results indicate that the proposed one-step approach substantially improves the performance of both LSE and WLSE methods, particularly for small sample sizes (e.g., $n \leq 100$).
Figure \ref{bias_mse_Rtheta_r1} facilitates a clearer comparison among the remaining methods by excluding LSE and WLSE. It is observed that the bias of the MLE decreases for both $\theta$ and $R$ as the sample size increases; a similar trend is generally evident for the one-step estimators.
In terms of MSE, the MLE demonstrates the most stable and robust performance among all methods for estimating $\theta$. Although the MSE values of the MPS, one-step LSE and one-step WLSE methods exhibit some fluctuations for $\theta$, the differences in MSE across all methods are relatively small for the reliability $R$. In particular, for large sample sizes, the MSE values of all methods become quite similar. For smaller samples, however, the MLE and MPS methods tend to yield lower MSE values, except in cases corresponding to small values of $\theta$.

As a summary of both simulation settings, based on all tables and figures, the MLE and MPS methods demonstrate more stable and efficient performance in terms of bias and MSE for nearly all parameters, with the exception of the copula parameter $\theta$ under the MPS method, across a range of dependence levels.
Although some irregularities are observed in the estimation of the copula parameter $\theta$ for certain methods, their impact on the estimation of the reliability parameter $R$ remains limited.

\begin{table}[ht]
\centering
\caption{
Estimates of model parameters, including $\theta$ and $R$, using ML, LS, WLS, and MPS, with one-step LS and WLS variants applied to $\theta$ and $R$, for $\theta=2$ and $3$.}
\label{table:estimates1}
\resizebox{\textwidth}{!}{
\begin{tabular}{cccccccccccccc}
  \hline
\multicolumn{2}{c}{} & \multicolumn{3}{c}{MLE} & \multicolumn{3}{c}{LSE} & \multicolumn{3}{c}{WLSE} & \multicolumn{3}{c}{MPS}  
  \\  \cmidrule(lr){3-5} \cmidrule(lr){6-8} \cmidrule(lr){9-11} \cmidrule(lr){12-14}
  n & Par & Estimates & Bias & MSE & Estimates & Bias & MSE & Estimates & Bias & MSE & Estimates & Bias & MSE \\ 
  \hline
  \multicolumn{2}{c}{} & \multicolumn{12}{c}{$\theta = 2, \; R = 0.76403$ } 
  \\  \hline 
25 & $a_1$ & 0.82384 & 0.07384 & 0.24654 & 0.85304 & 0.10304 & 0.36598 & 0.86239 & 0.11239 & 0.33547 & 0.78676 & 0.03676 & 0.21654 \\  
   & $b_1$ & 1.28578 & 0.03578 & 0.22095 & 1.15778 & -0.09222 & 0.35919 & 1.21596 & -0.03404 & 0.30685 & 1.09625 & -0.15375 & 0.23313 \\ 
   & $\lambda_1$ & 0.72582 & 0.12582 & 0.35425 & 0.80209 & 0.20209 & 0.74879 & 0.75014 & 0.15014 & 0.59117 & 0.71804 & 0.11804 & 0.38414 \\ 
   & $a_2$ & 1.82255 & -0.17745 & 0.99484 & 1.82063 & -0.17937 & 1.48284 & 1.88957 & -0.11043 & 1.37433 & 1.63983 & -0.36017 & 0.84788\\ 
   & $b_2$ & 1.40775 & -0.09225 & 0.16552 & 1.29932 & -0.20068 & 0.31884 & 1.36411 & -0.13589 & 0.25419 & 1.23405 & -0.26595 & 0.22064 \\ 
   & $\lambda_2$ & 0.61904 & 0.36904 & 0.68282 & 0.76071 & 0.51071 & 1.36943 & 0.67196 & 0.42196 & 1.03528 & 0.59664 & 0.34664 & 0.71366 \\ 
   & $\theta$ & 2.16301 & 0.16301 & 0.70367 & 2.22103 & 0.22103 & 4.41201 & 2.19761 & 0.19761 & 1.21987 & 2.41558 & 0.41558 & 1.88185\\ 
   & $R$ & 0.76967 & 0.00564 & 0.00494 & 0.75312 & -0.01091 & 0.00593 & 0.76097 & -0.00306 & 0.00564 & 0.74962 & -0.01441 & 0.00634\\ \hline
   \\
50 & $a_1$ & 0.80312 & 0.05312 & 0.14282 & 0.82363 & 0.07363 & 0.21632 & 0.83466 & 0.08466 & 0.19171 & 0.79387 & 0.04387 & 0.14193 \\ 
   & $b_1$ &  1.28646 & 0.03646 & 0.13066 & 1.22067 & -0.02933 & 0.19613 & 1.26855 & 0.01855 & 0.16049 & 1.18118 & -0.06882 & 0.13581\\ 
   & $\lambda_1$ & 0.65631 & 0.05631 & 0.20240 & 0.69326 & 0.09326 & 0.38193 & 0.65035 & 0.05035 & 0.28621 & 0.63972 & 0.03972 & 0.21738 \\ 
   & $a_2$ & 1.87521 & -0.12479 & 0.61931 & 1.86111 & -0.13889 & 0.85900 & 1.93118 & -0.06882 & 0.74834 & 1.78851 & -0.21149 & 0.56117  \\ 
   & $b_2$ & 1.45630 & -0.04370 & 0.09204 & 1.39224 & -0.10776 & 0.14987 & 1.44705 & -0.05295 & 0.11211 & 1.35948 & -0.14052 & 0.10720  \\ 
   & $\lambda_2$ &  0.45860 & 0.20860 & 0.29634 & 0.53547 & 0.28547 & 0.56765 & 0.45630 & 0.20630 & 0.37104 & 0.42855 & 0.17855 & 0.29358  \\ 
   & $\theta$ & 2.04767 & 0.04767 & 0.26675 & 2.05886 & 0.05886 & 1.78762 & 2.05890 & 0.05890 & 0.97736 & 2.18101 & 0.18101 & 1.38110 \\ 
   & $R $& 0.76750 & 0.00347 & 0.00247 & 0.75723 & -0.00680 & 0.00383 & 0.76279 & -0.00123 & 0.00321 & 0.75361 & -0.01042 & 0.00378\\ \hline
   \\
100 & $a_1$ & 0.80352 & 0.05352 & 0.09144 & 0.82574 & 0.07574 & 0.14571 & 0.82525 & 0.07525 & 0.11741 & 0.80706 & 0.05706 & 0.09772\\ 
   & $b_1$ & 1.28918 & 0.03918 & 0.07658 & 1.25955 & 0.00955 & 0.12060 & 1.28645 & 0.03645 & 0.09462 & 1.23365 & -0.01635 & 0.07809\\ 
   & $\lambda_1$ &  0.60608 & 0.00608 & 0.11276 & 0.61984 & 0.01984 & 0.21686 & 0.59675 & -0.00325 & 0.15211 & 0.58758 & -0.01242 & 0.12273  \\ 
   & $a_2$& 1.94625 & -0.05375 & 0.38487 & 1.90372 & -0.09628 & 0.54269 & 1.96193 & -0.03807 & 0.46282 & 1.92393 & -0.07607 & 0.35808\\ 
   & $b_2$ & 1.48088 & -0.01912 & 0.05269 & 1.43667 & -0.06333 & 0.08272 & 1.47383 & -0.02617 & 0.06320 & 1.43274 & -0.06726 & 0.05530 \\ 
   & $\lambda_2$ & 0.36296 & 0.11296 & 0.15249 & 0.42869 & 0.17869 & 0.29578 & 0.37522 & 0.12522 & 0.20007 & 0.32746 & 0.07746 & 0.14601 \\ 
   & $\theta$ & 2.01739 & 0.01739 & 0.14560 & 2.05251 & 0.05251 & 0.79594 & 2.04180 & 0.04180 & 0.39738 & 2.17539 & 0.17539 & 0.92575\\ 
   & $R$ & 0.76735 & 0.00333 & 0.00125 & 0.76344 & -0.00059 & 0.00250 & 0.76671 & 0.00268 & 0.00182 & 0.76227 & -0.00176 & 0.00275\\ 
   \hline
  \multicolumn{2}{c}{} & \multicolumn{12}{c}{$\theta = 3, \; R = 0.81533$ } 
  \\  \hline 
25 & $a_1$ &  0.88676 & 0.13676 & 0.25843 & 0.90333 & 0.15333 & 0.33618 & 0.90901 & 0.15901 & 0.31161 & 0.84547 & 0.09547 & 0.22351\\
   & $b_1$ & 1.37983 & 0.12983 & 0.20910 & 1.26981 & 0.01981 & 0.31118 & 1.31977 & 0.06977 & 0.27396 & 1.18471 & -0.06529 & 0.18831\\
   & $\lambda_1$ & 0.62212 & 0.02212 & 0.28865 & 0.67498 & 0.07498 & 0.53409 & 0.63983 & 0.03983 & 0.43758 & 0.61706 & 0.01706 & 0.32346 \\
   & $a_2$& 1.92494 & -0.07506 & 0.99051 & 1.84260 & -0.15740 & 1.67870 & 1.93236 & -0.06764 & 1.60597 & 1.72713 & -0.27287 & 0.80430 \\
   & $b_2$ & 1.48699 & -0.01301 & 0.13887 & 1.33394 & -0.16606 & 0.29307 & 1.41003 & -0.08997 & 0.23262 & 1.30695 & -0.19305 & 0.16807 \\ 
   & $\lambda_2$ &  0.55424 & 0.30424 & 0.56518 & 0.78745 & 0.53745 & 1.50814 & 0.68036 & 0.43036 & 1.15093 & 0.53185 & 0.28185 & 0.58762  \\
   & $\theta$ & 2.98397 & -0.01603 & 0.96030 & 3.15435 & 0.15435 & 7.26366 & 3.06753 & 0.06753 & 2.41253 & 3.30828 & 0.30828 & 3.21733  \\
   & $R$ &  0.81778 & 0.00245 & 0.00381 & 0.79971 & -0.01562 & 0.00568 & 0.80826 & -0.00707 & 0.00494 & 0.79363 & -0.02170 & 0.00576\\
\hline \\
50 & $a_1$ &  0.80618 & 0.05618 & 0.15766 & 0.82159 & 0.07159 & 0.22748 & 0.82941 & 0.07941 & 0.20920 & 0.79660 & 0.04660 & 0.15556 \\
   & $b_1$ & 1.27158 & 0.02158 & 0.11943 & 1.20396 & -0.04604 & 0.19391 & 1.24565 & -0.00435 & 0.15862 & 1.16616 & -0.08384 & 0.12778\\
   & $\lambda_1$ & 0.66145 & 0.06145 & 0.20339 & 0.70543 & 0.10543 & 0.38091 & 0.67000 & 0.07000 & 0.29224 & 0.64521 & 0.04521 & 0.22050 \\
   & $a_2$& 1.82450 & -0.17550 & 0.62188 & 1.79852 & -0.20148 & 0.88810 & 1.85553 & -0.14447 & 0.77237 & 1.74054 & -0.25946 & 0.57780  \\
   & $b_2$ & 1.43417 & -0.06583 & 0.08882 & 1.36341 & -0.13659 & 0.16989 & 1.41402 & -0.08598 & 0.12622 & 1.33686 & -0.16314 & 0.10814 \\ 
   & $\lambda_2$ & 0.48976 & 0.23976 & 0.32048 & 0.59584 & 0.34584 & 0.70338 & 0.52246 & 0.27246 & 0.49243 & 0.45955 & 0.20955 & 0.31665  \\
   & $\theta$ & 3.02447 & 0.02447 & 0.49734 & 3.16190 & 0.16190 & 4.47813 & 3.06751 & 0.06751 & 1.46462 & 3.21952 & 0.21952 & 1.56527  \\
   & $R$ & 0.81345 & -0.00188 & 0.00192 & 0.80165 & -0.01367 & 0.00392 & 0.80805 & -0.00728 & 0.00283 & 0.79874 & -0.01659 & 0.00336\\
\hline \\
100 & $a_1$ & 0.78744 & 0.03744 & 0.09116 & 0.80816 & 0.05816 & 0.15184 & 0.80801 & 0.05801 & 0.12637 & 0.79142 & 0.04142 & 0.09580 \\
   & $b_1$ & 1.26864 & 0.01864 & 0.06907 & 1.23344 & -0.01656 & 0.11976 & 1.26046 & 0.01046 & 0.09116 & 1.21418 & -0.03582 & 0.07260  \\
   & $\lambda_1$ & 0.62438 & 0.02438 & 0.11095 & 0.64274 & 0.04274 & 0.21891 & 0.62040 & 0.02040 & 0.15609 & 0.60460 & 0.00460 & 0.11918\\
   & $a_2$&   1.97185 & -0.02815 & 0.37177 & 1.91615 & -0.08385 & 0.57839 & 1.97466 & -0.02534 & 0.47043 & 1.95150 & -0.04850 & 0.34087 \\
   & $b_2$ & 1.48489 & -0.01511 & 0.04368 & 1.43548 & -0.06452 & 0.08202 & 1.47280 & -0.02720 & 0.05819 & 1.43760 & -0.06240 & 0.04542 \\ 
   & $\lambda_2$ & 0.34254 & 0.09254 & 0.13395 & 0.42994 & 0.17994 & 0.30830 & 0.36739 & 0.11739 & 0.19844 & 0.30532 & 0.05532 & 0.12766  \\
   & $\theta$ & 2.98175 & -0.01825 & 0.25720 & 3.04658 & 0.04658 & 1.37784 & 3.10721 & 0.10721 & 1.42797 & 3.10601 & 0.10601 & 1.26244  \\
   & $R$ & 0.81579 & 0.00047 & 0.00086 & 0.80994 & -0.00539 & 0.00256 & 0.81542 & 0.00009 & 0.00178 & 0.80596 & -0.00937 & 0.00222\\
\hline 
\end{tabular}
}
\end{table}

\begin{table}[ht]
\centering
\caption{
Estimates of model parameters, including $\theta$ and $R$, using ML, LS, WLS, and MPS, with one-step LS and WLS variants applied to $\theta$ and $R$, for $\theta=4$ and $6$.}
\label{table:estimates2}
\resizebox{\textwidth}{!}{
\begin{tabular}{cccccccccccccc}
  \hline
\multicolumn{2}{c}{} & \multicolumn{3}{c}{MLE} & \multicolumn{3}{c}{LSE} & \multicolumn{3}{c}{WLSE} & \multicolumn{3}{c}{MPS}  
  \\  \cmidrule(lr){3-5} \cmidrule(lr){6-8} \cmidrule(lr){9-11} \cmidrule(lr){12-14}
  n & Par & Estimates & Bias & MSE & Estimates & Bias & MSE & Estimates & Bias & MSE & Estimates & Bias & MSE \\ 
  \hline
  \multicolumn{2}{c}{} & \multicolumn{12}{c}{$\theta = 4, \; R = 0.85699$ } 
  \\  \hline 
  25 & $a_1$ & 0.88512 & 0.13512 & 0.26517 & 0.92348 & 0.17348 & 0.37785 & 0.93069 & 0.18069 & 0.35392 & 0.84279 & 0.09279 & 0.22828  \\
   & $b_1$ & 1.37832 & 0.12832 & 0.21640 & 1.28045 & 0.03045 & 0.31979 & 1.33378 & 0.08378 & 0.29019 & 1.18193 & -0.06807 & 0.19727 \\
   & $\lambda_1$ &  0.62065 & 0.02065 & 0.29756 & 0.65142 & 0.05142 & 0.51240 & 0.61614 & 0.01614 & 0.42140 & 0.61895 & 0.01895 & 0.33710 \\
   & $a_2$& 1.88888 & -0.11112 & 0.95233 & 1.89008 & -0.10992 & 1.71633 & 1.95047 & -0.04953 & 1.55270 & 1.69267 & -0.30733 & 0.80534  \\
   & $b_2$ & 1.48517 & -0.01483 & 0.15027 & 1.35231 & -0.14769 & 0.29987 & 1.42311 & -0.07689 & 0.23901 & 1.30283 & -0.19717 & 0.17993 \\ 
   & $\lambda_2$ & 0.56439 & 0.31439 & 0.56443 & 0.75138 & 0.50138 & 1.39369 & 0.66036 & 0.41036 & 1.04514 & 0.54621 & 0.29621 & 0.58727 \\
   & $\theta$ & 3.84050 & -0.15950 & 1.42873 & 4.13560 & 0.13560 & 12.81192 & 4.05304 & 0.05304 & 3.29285 & 4.24240 & 0.24240 & 3.86023  \\
   & $R$ & 0.85192 & -0.00507 & 0.00338 & 0.83364 & -0.02335 & 0.00590 & 0.84475 & -0.01223 & 0.00468 & 0.82795 & -0.02903 & 0.00588 \\
\hline \\
50 & $a_1$ & 0.81933 & 0.06933 & 0.14665 & 0.84995 & 0.09995 & 0.22579 & 0.85271 & 0.10271 & 0.19931 & 0.80925 & 0.05925 & 0.14642 \\
   & $b_1$ & 1.29575 & 0.04575 & 0.12835 & 1.25279 & 0.00279 & 0.21048 & 1.28755 & 0.03755 & 0.16891 & 1.18944 & -0.06056 & 0.13031 \\
   & $\lambda_1$ &  0.63049 & 0.03049 & 0.19169 & 0.66302 & 0.06302 & 0.37665 & 0.63016 & 0.03016 & 0.28896 & 0.61511 & 0.01511 & 0.20830 \\
   & $a_2$& 1.87509 & -0.12491 & 0.64878 & 1.83957 & -0.16043 & 0.91465 & 1.90096 & -0.09904 & 0.79040 & 1.78275 & -0.21725 & 0.58692  \\
   & $b_2$ & 1.45023 & -0.04977 & 0.09904 & 1.38585 & -0.11415 & 0.16559 & 1.43554 & -0.06446 & 0.12753 & 1.35154 & -0.14846 & 0.11428\\ 
   & $\lambda_2$ & 0.46588 & 0.21588 & 0.29896 & 0.57536 & 0.32536 & 0.65262 & 0.50185 & 0.25185 & 0.46762 & 0.43822 & 0.18822 & 0.29732 \\
   & $\theta$ & 3.98010 & -0.01990 & 0.82994 & 4.18088 & 0.18088 & 5.99197 & 4.12344 & 0.12344 & 3.04707 & 4.19369 & 0.19369 & 1.77459  \\
   & $R$ & 0.85507 & -0.00192 & 0.00157 & 0.84468 & -0.01231 & 0.00397 & 0.85208 & -0.00490 & 0.00257 & 0.83943 & -0.01756 & 0.00305 \\
\hline \\
100 & $a_1$ & 0.79972 & 0.04972 & 0.09496 & 0.82350 & 0.07350 & 0.15892 & 0.82403 & 0.07403 & 0.13094 & 0.80472 & 0.05472 & 0.10029 \\
   & $b_1$ & 1.27666 & 0.02666 & 0.07460 & 1.24429 & -0.00571 & 0.13081 & 1.27430 & 0.02430 & 0.09885 & 1.22311 & -0.02689 & 0.07721 \\
   & $\lambda_1$ & 0.61086 & 0.01086 & 0.11164 & 0.63238 & 0.03238 & 0.23639 & 0.60563 & 0.00563 & 0.16364 & 0.58982 & -0.01018 & 0.12033 \\
   & $a_2$& 1.94769 & -0.05231 & 0.40086 & 1.90382 & -0.09618 & 0.58692 & 1.96073 & -0.03927 & 0.48493 & 1.92218 & -0.07782 & 0.37073  \\
   & $b_2$ & 1.47187 & -0.02813 & 0.04848 & 1.42846 & -0.07154 & 0.08363 & 1.46555 & -0.03445 & 0.06115 & 1.42335 & -0.07665 & 0.05187\\ 
   & $\lambda_2$ &  0.35853 & 0.10853 & 0.15021 & 0.43539 & 0.18539 & 0.29792 & 0.37612 & 0.12612 & 0.19753 & 0.32469 & 0.07469 & 0.14388  \\
   & $\theta$ & 3.96275 & -0.03725 & 0.37127 & 4.15154 & 0.15154 & 3.75233 & 4.08555 & 0.08555 & 2.53400 & 4.13429 & 0.13429 & 1.49656  \\
   & $R$ & 0.85576 & -0.00122 & 0.00080 & 0.84915 & -0.00783 & 0.00275 & 0.85466 & -0.00233 & 0.00161 & 0.84585 & -0.01113 & 0.00200 \\
\hline 
  \multicolumn{2}{c}{} & \multicolumn{12}{c}{$\theta = 6, \; R = 0.91303$ } 
  \\  \hline 
  25 & $a_1$ &  0.84851 & 0.09851 & 0.25894 & 0.86811 & 0.11811 & 0.36722 & 0.87651 & 0.12651 & 0.35144 & 0.80745 & 0.05745 & 0.22349\\
   & $b_1$ &  1.32573 & 0.07573 & 0.22387 & 1.19798 & -0.05202 & 0.38234 & 1.25155 & 0.00155 & 0.33243 & 1.13104 & -0.11896 & 0.22242 \\
   & $\lambda_1$ & 0.69303 & 0.09303 & 0.36997 & 0.80287 & 0.20287 & 0.83594 & 0.75499 & 0.15499 & 0.68022 & 0.68804 & 0.08804 & 0.40091\\
   & $a_2$&  1.84987 & -0.15013 & 1.13080 & 1.84193 & -0.15807 & 1.59002 & 1.90483 & -0.09517 & 1.47904 & 1.64927 & -0.35073 & 0.92941\\
   & $b_2$ &  1.43887 & -0.06113 & 0.16458 & 1.30920 & -0.19080 & 0.33742 & 1.37693 & -0.12307 & 0.26846 & 1.25637 & -0.24363 & 0.21181 \\ 
   & $\lambda_2$ &  0.63184 & 0.38184 & 0.68396 & 0.82327 & 0.57327 & 1.72766 & 0.72703 & 0.47703 & 1.32692 & 0.61510 & 0.36510 & 0.70722\\
   & $\theta$ & 5.83836 & -0.16164 & 2.96218 & 6.31070 & 0.31070 & 7.47907 & 6.46399 & 0.46399 & 7.64950 & 6.65406 & 0.65406 & 8.98438 \\
   & $R$ & 0.90760 & -0.00544 & 0.00165 & 0.89334 & -0.01970 & 0.00398 & 0.90298 & -0.01005 & 0.00307 & 0.88534 & -0.02769 & 0.00424 \\
\hline \\
50 & $a_1$ & 0.79791 & 0.04791 & 0.14709 & 0.82274 & 0.07274 & 0.23183 & 0.82825 & 0.07825 & 0.20231 & 0.78859 & 0.03859 & 0.14753\\
   & $b_1$ & 1.26491 & 0.01491 & 0.12009 & 1.19691 & -0.05309 & 0.22517 & 1.24351 & -0.00649 & 0.17552 & 1.15920 & -0.09080 & 0.12970\\
   & $\lambda_1$ & 0.67159 & 0.07159 & 0.20085 & 0.73209 & 0.13209 & 0.47142 & 0.68536 & 0.08536 & 0.33858 & 0.65528 & 0.05528 & 0.21633\\
   & $a_2$& 1.83207 & -0.16793 & 0.63421 & 1.82205 & -0.17795 & 0.94575 & 1.88749 & -0.11251 & 0.84032 & 1.75252 & -0.24748 & 0.59009 \\
   & $b_2$ & 1.42753 & -0.07247 & 0.09059 & 1.36704 & -0.13296 & 0.16314 & 1.41617 & -0.08383 & 0.12627 & 1.33136 & -0.16864 & 0.11211\\ 
   & $\lambda_2$ & 0.49822 & 0.24822 & 0.35858 & 0.58255 & 0.33255 & 0.62742 & 0.51457 & 0.26457 & 0.48217 & 0.46630 & 0.21630 & 0.36000  \\
   & $\theta$ &  5.95169 & -0.04831 & 1.74383 & 6.19410 & 0.19410 & 5.52095 & 6.25192 & 0.25192 & 3.66115 & 6.33663 & 0.33663 & 3.16663   \\
   & $R$ & 0.90743 & -0.00560 & 0.00084 & 0.89609 & -0.01695 & 0.00256 & 0.90537 & -0.00767 & 0.00172 & 0.89340 & -0.01963 & 0.00197 \\
\hline \\
100 & $a_1$ &  0.80372 & 0.05372 & 0.09478 & 0.83455 & 0.08455 & 0.16427 & 0.83060 & 0.08060 & 0.13338 & 0.80773 & 0.05773 & 0.09995 \\
   & $b_1$ &  1.27532 & 0.02532 & 0.07026 & 1.25048 & 0.00048 & 0.12924 & 1.27415 & 0.02415 & 0.09634 & 1.22103 & -0.02897 & 0.07258\\
   & $\lambda_1$ & 0.61290 & 0.01290 & 0.11171 & 0.62758 & 0.02758 & 0.23567 & 0.60615 & 0.00615 & 0.16610 & 0.59262 & -0.00738 & 0.11980 \\
   & $a_2$& 1.94674 & -0.05326 & 0.39293 & 1.88986 & -0.11014 & 0.57768 & 1.95639 & -0.04361 & 0.48330 & 1.92264 & -0.07736 & 0.36222  \\
   & $b_2$ & 1.46978 & -0.03022 & 0.05037 & 1.41861 & -0.08139 & 0.08373 & 1.45873 & -0.04127 & 0.06179 & 1.42183 & -0.07817 & 0.05349 \\ 
   & $\lambda_2$ &  0.36116 & 0.11116 & 0.14026 & 0.44525 & 0.19525 & 0.31227 & 0.38046 & 0.13046 & 0.20120 & 0.32618 & 0.07618 & 0.13314  \\
   & $\theta$ & 5.90616 & -0.09384 & 0.84209 & 6.09747 & 0.09747 & 4.89967 & 6.10912 & 0.10912 & 2.24771 & 6.17439 & 0.17439 & 1.64020 \\
   & $R$ & 0.91029 & -0.00274 & 0.00044 & 0.90149 & -0.01155 & 0.00214 & 0.90949 & -0.00354 & 0.00095 & 0.90269 & -0.01034 & 0.00103 \\
\hline 
\end{tabular}
}
\end{table}

\begin{table}[ht]
\centering
\caption{
$95\%$ confidence intervals of model parameters using ML, LS, WLS, and MPS for $\theta = 2$ and $3$.}
\label{table:interval1}
\resizebox*{\textwidth}{\textheight}{
\begin{tabular}{cccccccccccc}
  \hline
\multicolumn{2}{c}{} & \multicolumn{2}{c}{MLE(ACI)} & \multicolumn{2}{c}{MLE(boot)} & \multicolumn{2}{c}{LSE(boot)} & \multicolumn{2}{c}{WLSE(boot)} & \multicolumn{2}{c}{MPS(boot)}  
  \\  \cmidrule(lr){3-4} \cmidrule(lr){5-6} \cmidrule(lr){7-8} \cmidrule(lr){9-10}  \cmidrule(lr){11-12}
  n & Par & Length & CP & Length & CP  & Length & CP & Length & CP & Length & CP   \\ 
  \hline
  \multicolumn{2}{c}{} & \multicolumn{10}{c}{$\theta = 2, \; R = 0.76403$ } 
  \\  \hline 
  25 & $a_1$ & 2.23426 & 0.900 & 1.79845 & 1.000 & 2.03039 & 0.999 & 2.00253 & 0.999 & 1.47457 & 0.985 \\
   & $b_1$ & 2.24869 & 0.957 & 1.90290 & 1.000 & 2.32480 & 0.999 & 2.22820 & 1.000 & 1.58158 & 0.977\\
   & $\lambda_1$ & 2.18091 & 0.973 & 2.33430 & 1.000 & 3.10207 & 1.000 & 2.89809 & 1.000 & 2.09361 & 0.998\\
   & $a_2$&  5.42151 & 0.847 & 3.89613 & 0.993 & 4.70990 & 0.994 & 4.65269 & 0.995 & 2.74400 & 0.949\\
   & $b_2$ & 2.21468 & 0.954 & 1.72533 & 1.000 & 2.20666 & 0.996 & 2.08342 & 0.999 & 1.44001 & 0.874\\ 
   & $\lambda_2$ & 2.54806 & 0.961 & 2.89820 & 0.996 & 4.17857 & 1.000 & 3.80990 & 1.000 & 2.56037 & 0.978 \\
   & $\theta$ & 3.30002 & 0.944 & 3.22714 &0.953 &4.96954 & 0.951 & 4.59197 & 0.949 & 5.99517 & 0.953 \\
   & $R$ & 0.48465 & 0.987 &0.26865 & 0.931 & 0.30183 & 0.954 & 0.29149 & 0.946 & 0.29999 & 0.963\\
\hline \\
50 & $a_1$ & 1.71180 & 0.902 & 1.37145 & 0.990 & 1.58434 & 0.997 & 1.53030 & 0.998 & 1.22904 & 0.966 \\
   & $b_1$ & 1.58218 & 0.964 & 1.37048 & 0.997 & 1.73251 & 0.999 & 1.57087 & 1.000 & 1.22299 & 0.963 \\
   & $\lambda_1$ & 1.59334 & 0.970 & 1.67184 & 0.987 & 2.23974 & 1.000 & 1.97390 & 0.999 & 1.54696 & 0.970\\
   & $a_2$&  4.35841 &0.884  & 2.84936 & 0.987 & 3.38169 & 0.995 & 3.26525 & 0.996 & 2.34309 & 0.954\\
   & $b_2$ & 1.57259 &0.962 & 1.22688 & 0.989 & 1.58918 & 0.992 & 1.42330 & 0.992 & 1.09474 & 0.901  \\ 
   & $\lambda_2$ & 1.76209 &0.974 & 1.95209 & 0.987 & 2.78763 & 0.997 & 2.37095 & 0.997 & 1.77593 & 0.972\\
   & $\theta$ & 2.11248 &0.942 & 2.15014 & 0.966 & 3.76253 & 0.945 & 2.96261 & 0.948 & 3.93169 & 0.972\\
   & $R$ & 0.37668 &0.997 & 0.19342 & 0.939 & 0.23450 & 0.952 & 0.21546 & 0.946 & 0.23420 & 0.972\\
\hline \\
100 & $a_1$ & 1.26668 & 0.933 & 1.07121 & 0.973 & 1.28129 & 0.990 & 1.19773 & 0.9870& 1.01528 & 0.963\\
   & $b_1$ & 1.10323 & 0.966 & 1.00710 & 0.974 & 1.28477 & 0.995 & 1.13187 & 0.995 & 0.94513 & 0.964\\
   & $\lambda_1$ &1.19909 & 0.966 & 1.22943 & 0.978 & 1.64162 & 0.991 & 1.41068 & 0.985 & 1.16774 & 0.967\\
   & $a_2$& 3.51364 & 0.911 & 2.23125 & 0.976 & 2.67751 & 0.982 & 2.50524 & 0.981 & 1.98324 & 0.972\\
   & $b_2$ & 1.09711 &  0.970 & 0.87973 & 0.988 & 1.14439 & 0.977 & 0.99845 & 0.988 & 0.81282 & 0.929\\ 
   & $\lambda_2$ & 1.25106 & 0.970 & 1.35479 & 0.972 & 1.93441 & 0.980& 1.59353 & 0.975 & 1.24279 & 0.974\\
   & $\theta$ & 1.42364 & 0.940 & 1.49590 & 0.959 & 3.22867 & 0.944 & 2.18190 & 0.949 & 3.78091 & 0.959\\
   & $R$ &0.27469 &  0.998 &0.13827 & 0.943 & 0.19084 & 0.950 & 0.16235 & 0.941 & 0.20351 & 0.967 \\
\hline 
  \multicolumn{2}{c}{} & \multicolumn{10}{c}{$\theta = 3, \; R = 0.81533$ } 
  \\  \hline 
  25 & $a_1$ & 2.41340 &0.941 & 1.82432 & 1.000 & 2.05360 & 1.000 & 2.02891 & 1.000 & 1.50320 & 0.996\\
   & $b_1$ & 2.38012 & 0.989 & 1.91722 & 1.000 & 2.36051 & 1.000 & 2.26009 & 1.000 & 1.60774 & 0.990 \\
   & $\lambda_1$ & 2.14335 & 0.995 & 2.28503 & 1.000 & 3.08809 & 1.000 & 2.88161 & 1.000 & 2.06138 & 0.999 \\
   & $a_2$& 5.86725 & 0.893 & 4.06129 & 0.998 & 4.93348 & 0.997 & 4.90617 & 0.997 & 2.84841 & 0.971 \\
   & $b_2$ & 2.35293 &0.981 & 1.77343 & 0.998 & 2.29059 & 0.995 & 2.17154 & 0.997 & 1.48762 & 0.912\\ 
   & $\lambda_2$ &  2.59576 & 0.985 & 2.93751 & 0.997 & 4.32373 & 1.000 & 3.95297 & 1.000 & 2.60651 & 0.993 \\
   & $\theta$ & 4.53336 &0.946 & 4.04618 & 0.969 & 6.86623 & 0.937 & 6.16027 & 0.956 & 7.90897 & 0.947\\
   & $R$ & 0.45858& 0.991 & 0.23999 & 0.942 & 0.28756 & 0.957 & 0.27159 & 0.958 & 0.28609 & 0.968\\
\hline \\
50 & $a_1$ & 1.70542 & 0.908 & 1.37366 & 0.993 & 1.58415 & 0.999 & 1.52604 & 0.997 & 1.22939 & 0.973\\
   & $b_1$ & 1.56301 & 0.967 &  1.36786 & 0.997 & 1.72799 & 0.998 & 1.56990 & 0.999 & 1.21868 & 0.964\\
   & $\lambda_1$ & 1.59020 & 0.977 &  1.67303 & 0.991 & 2.24740 & 1.000 & 1.98443 & 1.000 & 1.54664 & 0.976\\
   & $a_2$& 4.21695 & 0.874 & 2.84788 & 0.988 & 3.37481 & 0.996 & 3.25362 & 0.997 & 2.34035 & 0.953\\
   & $b_2$ & 1.55620 & 0.960 & 1.22925 & 0.995 & 1.59502 & 0.988 & 1.43385 & 0.993 & 1.09798 & 0.906\\ 
   & $\lambda_2$ & 1.78623 & 0.967 &1.97216 & 0.983 & 2.82977 & 0.998 & 2.42378 & 0.997 & 1.79570 & 0.967 \\
   & $\theta$ & 2.99648 & 0.938 & 2.84161 & 0.966 & 5.39033 & 0.940 & 4.09714 & 0.958 & 4.63774 & 0.959\\
   & $R$ & 0.38795 & 0.995 & 0.17480 & 0.958 & 0.23198 & 0.948 & 0.20623 & 0.961 & 0.21714 & 0.959\\
\hline \\
100 & $a_1$ & 1.23217 & 0.919 & 1.06189 & 0.974 & 1.26773 & 0.987 & 1.18309 & 0.985 & 1.00816 & 0.958\\
   & $b_1$ & 1.09057 & 0.970 & 1.00512 & 0.976 & 1.28161 & 0.996 & 1.13003 & 0.997 & 0.94342 & 0.963\\
   & $\lambda_1$ & 1.20433 &  0.968 & 1.23404 & 0.978 & 1.64834 & 0.990 & 1.41854 & 0.991 & 1.17344 & 0.966 \\
   & $a_2$& 3.53577 & 0.926 & 2.23274 & 0.986 & 2.67860 & 0.992 & 2.49955 & 0.988 & 1.98191 & 0.979\\
   & $b_2$ & 1.09406 & 0.977 &  0.87450 & 0.989 & 1.14401 & 0.991 & 0.99457 & 0.993 & 0.80767 & 0.956\\
   & $\lambda_2$ & 1.22938 & 0.981 & 1.33692 & 0.978 & 1.93805 & 0.983 & 1.58327 & 0.982 & 1.22211 & 0.980\\
   & $\theta$ & 1.97361 & 0.946 &  1.98514 & 0.965 & 4.62583 & 0.951 & 3.09119 & 0.933 & 3.91925 & 0.971\\
   & $R$ & 0.30016 & 0.990 & 0.12352 & 0.968 & 0.19473 & 0.955 & 0.15710 & 0.944 & 0.18287 & 0.973\\
\hline 
\end{tabular}
}
\end{table}

\begin{table}[ht]
\centering
\caption{$95\%$ confidence intervals of model parameters using ML, LS, WLS, and MPS for $\theta = 4$ and $6$.}
\label{table:interval2}
\resizebox*{\textwidth}{\textheight}{
\begin{tabular}{cccccccccccc}
  \hline
\multicolumn{2}{c}{} & \multicolumn{2}{c}{MLE(ACI)} & \multicolumn{2}{c}{MLE(boot)} & \multicolumn{2}{c}{LSE(boot)} & \multicolumn{2}{c}{WLSE(boot)} & \multicolumn{2}{c}{MPS(boot)}  
  \\  \cmidrule(lr){3-4} \cmidrule(lr){5-6} \cmidrule(lr){7-8} \cmidrule(lr){9-10}  \cmidrule(lr){11-12}
  n & Par & Length & CP & Length & CP  & Length & CP & Length & CP & Length & CP   \\ 
  \hline
  \multicolumn{2}{c}{} & \multicolumn{10}{c}{$\theta = 4, \; R = 0.85699$ } 
  \\  \hline 
  25 & $a_1$ & 2.40767 & 0.932 & 1.80071 & 1.000 & 2.05655 & 1.000 & 2.02458 & 1.000 & 1.48599 & 0.995 \\
   & $b_1$ & 2.37814 & 0.986 &  1.91669 & 1.000 & 2.35289 & 1.000 & 2.25924 & 1.000 & 1.60840 & 0.994 \\
   & $\lambda_1$ & 2.13424 & 0.994 & 2.26807 & 1.000 & 3.06947 & 1.000 & 2.85869 & 1.000 & 2.04541 & 0.997 \\
   & $a_2$& 5.75474 & 0.894 & 4.01037 & 0.998 & 4.94468 & 0.997 & 4.89386 & 0.999 & 2.82146 & 0.958 \\
   & $b_2$ & 2.36097 & 0.980 &  1.77779 & 0.998 & 2.29538 & 0.994 & 2.17746 & 0.997 & 1.49227 & 0.921 \\ 
   & $\lambda_2$ & 2.60669 & 0.983 & 2.93227 & 0.999 & 4.30106 & 1.000 & 3.92940 & 1.000 & 2.60835 & 0.993 \\
   & $\theta$ & 6.00304 & 0.927 & 4.90028 & 0.956 & 8.67222 & 0.942 & 7.97160 & 0.957 & 10.2367 & 0.956\\
   & $R$ & 0.42876 & 0.987 & 0.21590 & 0.940 & 0.27230 & 0.947 & 0.24990 & 0.949 & 0.27168 & 0.943 \\
\hline \\
50 & $a_1$ & 1.73461 & 0.929 &  1.37339 & 0.991 & 1.59697 & 1.000 & 1.53719 & 1.000 & 1.23009 & 0.978 \\
   & $b_1$ & 1.58161 & 0.976 & 1.36204 & 0.999 & 1.74137 & 1.000 & 1.57321 & 1.000 & 1.21609 & 0.975 \\
   & $\lambda_1$ & 1.56846 & 0.982 & 1.64887 & 0.995 & 2.23911 & 1.000 & 1.96889 & 1.000 & 1.52561 & 0.985 \\
   & $a_2$& 4.36932 &  0.877 & 2.86729 & 0.986 & 3.44201 & 0.996 & 3.31395 & 0.997 & 2.35529 & 0.953 \\
   & $b_2$ & 1.57021 & 0.962 & 1.22713 & 0.996 & 1.61046 & 0.993 & 1.44260 & 0.996 & 1.09635 & 0.897\\ 
   & $\lambda_2$ & 1.77039 & 0.974 & 1.95970 & 0.981 & 2.86081 & 0.999 & 2.43547 & 0.998 & 1.78711 & 0.971 \\
   & $\theta$ & 4.10003 &  0.944 &  3.52250 & 0.950 & 6.78747 & 0.933 & 5.28766 & 0.940 & 5.42014 & 0.953 \\
   & $R$ & 0.34888 & 0.998 & 0.15370 & 0.950 & 0.21764 & 0.939 & 0.18841 & 0.948 & 0.19922 & 0.933 \\
\hline \\
100 & $a_1$ & 1.24966 & 0.929 & 1.06390 & 0.971 & 1.26910 & 0.982 & 1.18837 & 0.981 & 1.00910 & 0.956 \\
   & $b_1$ & 1.09065 & 0.963 & 0.99912 & 0.972 & 1.27616 & 0.995 & 1.12519 & 0.993 & 0.93660 & 0.960\\
   & $\lambda_1$ & 1.18933 & 0.960 & 1.22140 & 0.978 & 1.63913 & 0.983& 1.40751 & 0.986 & 1.15820 & 0.971\\
   & $a_2$& 3.47567 & 0.915 & 2.20995 & 0.972 & 2.67550 & 0.983 & 2.49312 & 0.979 & 1.96396 & 0.958\\
   & $b_2$ & 1.08558 & 0.961 &  0.87256 & 0.982 & 1.14280 & 0.980 & 0.99349 & 0.985 & 0.80749 & 0.931\\ 
   & $\lambda_2$ & 1.23530 & 0.973 & 1.33788 & 0.968 & 1.94354 & 0.988 & 1.58990 & 0.980 & 1.22938 & 0.972\\
   & $\theta$ & 2.62294 & 0.942 &  2.48754 & 0.951 & 5.90658 & 0.942 & 3.90658 & 0.957 & 4.25616 & 0.961\\
   & $R$ & 0.28483 & 0.999 & 0.10850 & 0.952 & 0.18530 & 0.940 & 0.14611 & 0.938 & 0.16144 & 0.945\\
\hline 
  \multicolumn{2}{c}{} & \multicolumn{10}{c}{$\theta = 6, \; R = 0.91303$ } 
  \\  \hline 
  25 & $a_1$ & 2.29970 &0.910 & 1.80507 & 1.000 & 2.05215 & 1.000 & 2.02139 & 0.999 & 1.47917 & 0.985 \\
   & $b_1$ & 2.30402 & 0.968  & 1.91717 & 1.000 & 2.38471 & 1.000 & 2.28451 & 1.000 & 1.59793 & 0.983\\
   & $\lambda_1$ & 2.15740 & 0.973  & 2.32169 & 1.000 & 3.16856 & 1.000 & 2.95015 & 1.000 & 2.08695 & 0.993 \\
   & $a_2$& 5.65131 & 0.861 & 4.03684 & 0.996 & 4.89759 & 0.996 & 4.82634 & 0.999 & 2.81943 & 0.949\\
   & $b_2$ & 2.29967 &0.966 & 1.77211 & 0.997 & 2.27639 & 0.995 & 2.15274 & 0.994 & 1.48106 & 0.891  \\ 
   & $\lambda_2$ & 2.65446  & 0.973 & 2.98316 & 0.999 & 4.31596 & 1.000 & 3.94212 & 1.000 & 2.65002 & 0.980 \\
   & $\theta$ & 10.5052 &  0.945 & 6.88655 & 0.952 & 13.0018 & 0.961 & 12.7467 & 0.952 & 17.4187 & 0.973\\
   & $R$ & 0.34392 & 0.994 & 0.17079 & 0.975 & 0.23136 & 0.975 & 0.20270 & 0.971 & 0.23989 & 0.930 \\
\hline \\
50 & $a_1$ & 1.68620 & 0.911 & 1.37798 & 0.992 & 1.58571 & 0.997 & 1.52831 & 0.996 & 1.23208 & 0.973 \\
   & $b_1$ & 1.55951 & 0.965 & 1.37171 & 0.998 & 1.73661 & 1.000 & 1.57470 & 1.000 & 1.22179 & 0.967\\
   & $\lambda_1$ & 1.60095 &0.978& 1.68273 & 0.991 & 2.26948 & 0.998 & 1.99821 & 0.996 & 1.55794 & 0.983  \\
   & $a_2$& 4.22890 & 0.877 & 2.85022 & 0.982 & 3.40616 & 0.991 & 3.28514 & 0.989 & 2.33574 & 0.943 \\
   & $b_2$ & 1.55115 & 0.957 & 1.22874 & 0.990 & 1.59703 & 0.989 & 1.43267 & 0.994 & 1.09528 & 0.892   \\ 
   & $\lambda_2$ & 1.78538 & 0.957 & 1.97374 & 0.973 & 2.83820 & 0.994 & 2.42225 & 0.990 & 1.79216 & 0.958 \\
   & $\theta$ & 6.98264 & 0.961 & 4.93063 & 0.945& 9.12476 & 0.963 & 7.62735 & 0.943 & 7.44493 & 0.943\\
   & $R$ & 0.28027 & 0.999 & 0.12116 & 0.976 & 0.18814 & 0.975 & 0.15363 & 0.967 & 0.16611 & 0.927 \\
\hline \\
100 & $a_1$ & 1.25652 & 0.928 & 1.07374 & 0.978 & 1.28083 & 0.987 & 1.19605 & 0.983 & 1.01839 & 0.957  \\
   & $b_1$ & 1.09022 & 0.971 & 0.99924 & 0.986 & 1.27484 & 0.996 & 1.12289 & 0.997 & 0.93891 & 0.960 \\
   & $\lambda_1$ & 1.19571 & 0.960  & 1.22753 & 0.977 & 1.64329 & 0.985 & 1.40930 & 0.983 & 1.16634 & 0.964  \\
   & $a_2$&  3.48995 &  0.922 & 2.23036 & 0.982 & 2.67893 & 0.986 & 2.50009 & 0.986 & 1.98406 & 0.971 \\
   & $b_2$ & 1.08566 & 0.970  & 0.87478 & 0.985 & 1.14133 & 0.988 & 0.99197 & 0.989 & 0.80923 & 0.928 \\ 
   & $\lambda_2$ & 1.24408 & 0.974 & 1.34735 & 0.972 & 1.94869 & 0.987 & 1.59405 & 0.984 & 1.23711 & 0.976 \\
   & $\theta$ & 4.26265 & 0.949 & 3.47720 & 0.947 & 7.85549 & 0.946 & 5.43292 & 0.941 & 5.12717 & 0.952 \\
   & $R$ & 0.21884 & 0.998 & 0.08438 & 0.960 & 0.15655 & 0.956 & 0.11704 & 0.954 & 0.12256 & 0.923\\
\hline 
\end{tabular}
}
\end{table}

\section{Real data application}\label{S:real_data}

In the reliability literature, dam or reservoir occupancy data were first utilized by \cite{KizilaslanNadar2018} to illustrate an application of multicomponent stress–strength models based on bivariate distribution. Since then, this line of research has expanded, with various studies employing such data under different modeling frameworks and scenarios. However, as noted in the Introduction, the literature on dependent stress–strength reliability remains relatively limited. Moreover, existing studies often lack carefully constructed real-data applications that adequately reflect both the underlying dependence structure and a meaningful practical scenario.
In this context, well-motivated real-life applications are essential not only for validating theoretical developments but also for demonstrating their practical relevance. Therefore, in this section, we construct a realistic and interpretable scenario to apply the proposed dependent stress–strength reliability model.

Istanbul, the largest city in T\"{u}rkiye, has a population of approximately $16$ million according to the $2025$ address-based registration records reported by the Turkish Statistical Institute \url{https://veriportali.tuik.gov.tr/en/press/53899}. It is one of the most populous metropolitan areas in the world and is uniquely located across both Europe and Asia. A significant proportion of the population (approximately $66\%$) resides on the European side, while the remainder lives on the Asian (Anatolian) side.

The largest dam supplying the city, Omerli Dam, is located on the Anatolian side and is supported by additional water inflow from nearby river systems. Meanwhile, water demand on the European side is partially met through an undersea tunnel, enabling water transfer from the Anatolian side. In recent years, drought risk has become a critical concern in Istanbul, particularly during late summer and autumn, when dam occupancy levels can drop significantly. Under such conditions, increased water transfer is required, leading to higher utilization of the transmission infrastructure, elevated energy costs, and increased operational risk.
Consequently, monitoring and comparing the occupancy levels of major dams on both sides of the city is of practical importance. Given that water demand patterns across the two regions are broadly similar, it is reasonable to expect a natural dependence between their occupancy levels. This dependence structure will be formally assessed using statistical methods.

For this purpose, we obtain data from the open data platform of the Istanbul Metropolitan Municipality \url{https://data.ibb.gov.tr/en/}. Similar datasets have been used in recent studies such as \cite{akgul2023estimation} and \cite{curan2025statistical} for the stress-strength reliability modeling. The dataset consists of daily occupancy rates of Istanbul’s dams, retrieved in March $2026$ from \href{https://data.ibb.gov.tr/en/dataset/istanbul-barajlari-gunluk-doluluk-oranlari/resource/af0b3902-cfd9-4096-85f7-e2c3017e4f21}{Istanbul Dams Daily Occupancy Rates}.
In this study, we focus on the two largest dams: Omerli (Anatolian side) and Terkos (European side). The data span the period from late October $2000$ to mid-February $2024$.
To reduce variability and focus on seasonal effects, we compute monthly average occupancy rates for the period September–December of each year. This results in a total of $95$ paired observations for both variables. 

In the stress–strength framework, the Terkos dam is treated as the strength variable $X$, while the Omerli dam is considered as the stress variable $Y$.
Under this formulation, the stress–strength reliability $R=P(X>Y)$ represents the probability that the occupancy level of the Terkos dam (European side) exceeds that of the Omerli dam (Anatolian side). This probability quantifies the likelihood that local water availability on the European side is sufficient relative to the corresponding level on the Anatolian side.
In practical terms, when the Terkos level falls below that of Omerli, additional water transfer to the European side may be required. A higher value of $R$ indicates a more favorable scenario, implying a reduced need for water transfer and lower operational costs, whereas lower values of $R$ suggest increased reliance on inter-regional water transfer and, consequently, higher system costs and risks.

To model the dependence between the strength and stress variables, we first applied the Pearson correlation test to assess linear dependence. The correlation coefficient between data $\bm{X}$ (Terkos dam) and $\bm{Y}$ (Omerli dam) was calculated as $0.49432$, with a corresponding t-test statistic of $5.484$ and a p-value of $0$, indicating a statistically significant positive linear association. 
Subsequently, we conducted goodness-of-fit tests for copula selection, considering the Clayton, Gumbel, and Frank copulas. The results are presented in Table \ref{table:gof_copula}, which show that all three copulas provide an adequate fit for characterizing the dependence structure between $X$ and $Y$. Given that the Clayton copula is capable of capturing lower tail dependence and that the dam occupancy rates frequently exhibit simultaneous low values on a monthly basis, it was deemed appropriate to model the data using the Clayton copula in our analysis.

Following the copula assessment, we evaluated the marginal distributions of $\bm X$ and $\bm Y$ using the Kolmogorov–Smirnov (K–S), Anderson–Darling (A–D), and Cramér–von Mises (C–VM) goodness-of-fit tests. These tests were performed to assess whether the data are consistent with the MWD or the two-parameter Weibull distribution. The test statistics and corresponding p-values (reported in parentheses), calculated based on the MLE of the unknown parameters, are presented in Table \ref{table:gof_data}, alongside the Akaike information criterion (AIC) and Bayesian information criterion (BIC) values. Based on these results, it is concluded that, although the Weibull distribution provides a slightly better fit for $\bm X$ alone, the MWD adequately fits both datasets and is therefore used in the subsequent stress–strength reliability analysis.

Finally, all proposed methods are applied to the real dataset to obtain point estimates and corresponding $95\%$ bootstrap confidence intervals (based on $B = 1000$ replications) for both the model parameters and the stress-strength reliability $R$. The results are presented in Table \ref{table:data_analysis}.
It is observed that the estimated values of the reliability $R$ are close to $0.50$ across all methods, although some variation is present in the estimates of some remaining parameters.
Based on the application scenario described above, a reliability value around $0.50$ suggests a relatively balanced situation between the two systems. This indicates that, at present, there may be no immediate need for substantial water transfer from the Anatolian side to the European side. However, given the inherent variability and potential risk of decline in water levels, the system should be monitored closely by the relevant authorities.

\begin{table}[h!]
\centering
\caption{Goodness-of-fit test results for copula.}
\label{table:gof_copula}
\begin{tabular}{cccc}
  \hline
 Copula & $\hat{\theta}$ & Test statistic & p-value \\ 
  \hline
Clayton & 1.25912 & 0.03570 & 0.10557 \\ 
  Gumbel & 1.60906 & 0.02573 & 0.15204 \\ 
  Frank & 4.06336 & 0.02495 & 0.13593 \\ 
   \hline
\end{tabular}
\end{table}

\begin{table}[h!]
\centering
\caption{MLEs of model parameters, goodness-of-fit test results, AIC and BIC values for the $\bm X$ and $\bm Y$ data sets.}
\label{table:gof_data}
\begin{tabular}{cccc}
  \hline
 Data &Parameter/ Criteria & Modified Weibull D. & Weibull D.\\ 
  \hline
$\bm{X}$ & $\hat{a}_1$ & 0.98577 & 6.92779  \\ 
       & $\hat{b}_1$ & 2.66112 &  3.91084 \\ 
       & $\hat{\lambda}_1$ & 2.11279 &  -  \\ 
       & KS & 0.09901 (0.28991) & 0.08717 (0.44108) \\ 
       & A-D & 1.22755 (0.25720) & 1.21529 (0.26170) \\ 
       & C-VM & 0.20246 (0.26299) & 0.19608 (0.27545) \\ 
       & AIC & -73.69488 & -74.53015  \\ 
       & BIC & -66.03325 &  -69.42240 \\ 
       \\
$\bm{Y}$ & $\hat{a}_2$ & 0.02428 &  6.02054 \\ 
       & $\hat{b}_2$ & 0.45908 &  3.52384  \\ 
       & $\hat{\lambda}_2$ & 6.33059 & -   \\ 
       & KS & 0.06459 (0.79867) &0.12571 (0.09096) \\ 
       & A-D & 0.59443 (0.65284) &2.72945 (0.03775) \\ 
       & C-VM & 0.07151 (0.74348) &0.37953 (0.08170) \\ 
       & AIC & -68.45257 & -49.26487  \\ 
       & BIC & -60.79094 & -44.15712 \\ 
   \hline
\end{tabular}
\end{table}

\begin{table}[ht]
\centering
\caption{Point estimates and $95\%$ bootstrap confidence intervals for real data under MLE, LSE, WLSE, and MPS.}
\label{table:data_analysis}
\resizebox{\textwidth}{!}{
\begin{tabular}{ccccccccc}
  \hline
\multicolumn{1}{c}{}  & \multicolumn{2}{c}{MLE} & \multicolumn{2}{c}{LSE} & \multicolumn{2}{c}{WLSE} & \multicolumn{2}{c}{MPS}  \\  
  \cmidrule(lr){2-3} \cmidrule(lr){4-5} \cmidrule(lr){6-7} \cmidrule(lr){8-9}  
   Par & Estimate & Interval & Estimate & Interval  & Estimate & Interval & Estimate & Interval   \\  
   \hline 
    $a_1$ & 0.98542 & (0.04928, 8.52758) & 0.07196 & (0.00191, 18.53088) & 0.04261 & (0.00237, 15.08527)& 0.68997 & (0.03377, 7.02578)\\
    $b_1$ & 2.66095 & (0.99417, 4.40638)& 1.72091 & (0.00001, 5.86277) & 1.29079 & (0.00001, 5.47741)& 2.31111 & (0.57907, 4.00116)\\
    $\lambda_1$ & 2.11322 &(0.00001, 5.63775)  & 5.71667 & (0.00001, 10.21649) & 6.20629 & (0.00001, 9.74951 )& 2.39000 & (0.00001, 5.87764)\\
    $a_2$& 0.02428  & (0.00937, 1.14113) & 0.01166 & (0.00372, 6.14717) & 0.01154 & (0.00421, 1.68287)&  0.02043& (0.00790, 0.44550\\
    $b_2$ & 0.45908 & (0.19583, 2 .90662) & 0.00001 &  (0.00001, 4.06822)& 0.00001 & (0.00001, 3.40848)&  0.26258& (0.00001, 2.17920)\\
    $\lambda_2$ & 6.33059  & (2.22795, 7.57024) & 7.06623 & (0.00001, 8.66182) & 7.10656 & (1.85330, 8.53702)& 6.43699 & (3.02716, 7.66614) \\
    $\theta$ & 0.50551 & (0.16689, 0.89215) & 1.09884 &  (0.48896, 2.28199)& 1.23631 & (0.64875, 2.08638)& 0.92547& (0.17318, 3.69228) \\
    $R$ & 0.50428 & (0.42256, 0.58432) &  0.49349& (0.39376, 0.57406)  & 0.49824&(0.40909, 0.58370)  & 0.51110& (0.41924, 0.60860|)\\
\hline \\
\end{tabular}
}
\end{table}

\section{Discussion and conclusion}\label{S:conclusion}

In this study, we developed a dependent stress–strength reliability model in which both stress and strength components follow modified Weibull distributions, with their dependence captured via a Clayton copula. The proposed seven-parameter framework extends classical Weibull-based models and provides a flexible structure for modeling system reliability under dependence.

Through extensive Monte Carlo simulations, we demonstrated that the proposed estimators—maximum likelihood (ML), least squares (LS), one-step LS, weighted least squares (WLS), one-step WLS, and maximum product of spacings (MPS)—yield accurate and reliable inference for both model parameters and the reliability $R$. Bootstrap confidence intervals were found to provide improved coverage, particularly for the copula parameter $\theta$ and $R$. Among the methods considered, MLE generally exhibits the best overall performance, especially in estimating $\theta$ and $R$, while MPS yields comparable performance for $R$. In addition, MLE and MPS show similarly strong performance in terms of bias and MSE for the marginal model parameters. The proposed one-step LSE and WLSE methods further improve upon their classical counterparts, particularly for small sample sizes.

The real-data application based on monthly occupancy levels of Istanbul’s two largest dams demonstrates the practical relevance of the proposed model. The Clayton copula effectively captures lower-tail dependence, and the resulting stress–strength reliability estimates provide meaningful insights into water transfer requirements, emphasizing the importance of incorporating dependence in operational risk assessment.

Overall, the proposed framework offers a flexible and effective tool for analyzing dependent stress–strength systems in engineering and environmental applications. Future research may consider extensions to more complex settings, such as multicomponent systems, alternative copula families, and different lifetime distributions, to further enhance the applicability of the model in real-world scenarios.

\clearpage
\section*{Software and reproducibility}
An R package implementing the proposed modeling framework and associated inference algorithms is publicly available at our GitHub repository: \url{https://github.com/fatihki/SSReliabilityClaytonMWD}.
It can be installed via \texttt{devtools::install\_github("fatihki/SSReliabilityClaytonMWD")}.

\section*{Data Availability}
The real datasets analyzed during the current study are available in the same GitHub repository.

\appendix
\section{Appendix}\label{appendix}
\numberwithin{equation}{section} 
\setcounter{equation}{0}         
\numberwithin{figure}{section} 
\setcounter{figure}{0} 

\subsection{Quadratic approximation of $C_{\theta}(u,v)$ in $Q_3(\theta)$ and $Q_3^W(\theta)$ } \label{A:approx_LS_WLS}

We apply a second-order order Taylor expansion for $C_{\theta} \equiv C_{\theta}(u,v)$ around $\theta_0$ for the LS and WLS estimates of $\theta$ parameter to be able to getting more robust estimate. 
Let $C_{\theta} = (S(\theta))^{-1/\theta}$ where $S(\theta) = (u^{-\theta} +  v^{-\theta} -1)$. Then, the first and second derivatives of $C_{\theta}$ are obtained as 
\begin{equation*}
    \dot{C}_{\theta} \equiv \frac{\partial C_{\theta}}{\partial \theta} = C_{\theta} \; d_1, \; \; 
    d_1 = \frac{\log(S(\theta))}{\theta^2} + \frac{1}{\theta S(\theta)} \frac{\partial S}{ \partial \theta}, 
\end{equation*}
and 
\begin{equation*}
     \ddot{C}_{\theta} \equiv \frac{\partial^2 C_{\theta}}{\partial \theta^2} = C_{\theta} \bigg( d_1^{2} + \frac{\partial d_1}{\partial \theta} \bigg), \; \frac{\partial d_1}{\partial \theta} = -\frac{2\log(S(\theta))}{\theta^3} - \frac{1}{\theta S^2(\theta)} \bigg( \frac{\partial S(\theta)}{\partial \theta} \bigg) ^2  
     + \frac{1}{\theta S(\theta)} \frac{\partial^2 S(\theta)}{\partial \theta^2}.
\end{equation*}
Then, the approximation is given by
\begin{equation*}
    C_{\theta}(u,v) \approx C_{\theta_0} + (\theta - \theta_0) \dot{C}_{\theta_0} + \frac{1}{2} (\theta - \theta_0)^2 \; \ddot{C}_{\theta_0}.
\end{equation*}

When it is implementing in the objective function $Q_3(\theta)$ in (\ref{LS_eq_theta}) of minimizing process for the LSE method, it becomes as 
\begin{align*}
    Q_3(\theta)  &= \sum_{i=1}^n \bigg( C_{\theta}(u_i,v_i;\theta) - \hat{H}_n(x_i,y_i) \bigg)^2 \\
    &\approx \sum_{i=1}^n \bigg(  C_{\theta_0} + (\theta - \theta_0) \dot{C}_{\theta_0} + \frac{1}{2} (\theta - \theta_0)^2 \; \ddot{C}_{\theta_0} - \hat{H}_n \bigg)^2, \;\; \hat{H}_n \equiv \hat{H}_n(x_i,y_i) \\ 
    & = \sum_{i=1}^n (C_{\theta_0} - \hat{H}_n )^2 + 2(\theta - \theta_0) \sum_{i=1}^n \dot{C}_{\theta_0} (C_{\theta_0} - \hat{H}_n)  
    + (\theta - \theta_0)^2 \sum_{i=1}^n  \bigg( \dot{C}_{\theta_0} ^2 + (C_{\theta_0} - \hat{H}_n) \ddot{C}_{\theta_0}  \bigg) + O((\theta - \theta_0)^3).
\end{align*}
Then, it can be rewritten as 
\begin{equation}
    Q_3(\theta) \approx A + B(\theta - \theta_0) +  C (\theta - \theta_0)^2 + O((\theta - \theta_0)^3), \label{LS_theta_apprx}
\end{equation}
where 
\begin{align*}
    A = \sum_{i=1}^n (C_{\theta_0} - \hat{H}_n )^2, \; B = 2 \sum_{i=1}^n \dot{C}_{\theta_0} (C_{\theta_0} - \hat{H}_n) , \; 
    C = \sum_{i=1}^n  \bigg( \dot{C}_{\theta_0} ^2 + (C_{\theta_0} - \hat{H}_n) \ddot{C}_{\theta_0}  \bigg).
\end{align*}
The remainder term is $O((\theta - \theta_0)^3)$, which is negligible for $\theta$ close to $\theta_0$.
We use Kendall’s $\tau$ estimate $\hat{\theta}_{\tau}$ as an initial value, namely $\theta_0 = \hat{\theta}_{\tau}$.
Therefore, minimizing this quadratic approximation in (\ref{LS_theta_apprx}) leads to a one-step update:
\begin{equation}
    \hat{\theta}_{LSE, one-step} = \hat{\theta}_{\tau} - \frac{B}{2C}. \label{LSE_one_step}
\end{equation}

Similarly, a weighted version is obtained for the WLSE method by including weights $w_i$ in the sums B and C as 
\begin{equation}
    \hat{\theta}_{WLSE, one-step} = \hat{\theta}_{\tau}- \frac{B^*}{2C^*},
    \label{WLSE_one_step}
\end{equation}
where 
\begin{equation*}
     B^* = 2 \sum_{i=1}^n w_i \dot{C}_{\hat{\theta}_{\tau}} (C_{\hat{\theta}_{\tau}} - \hat{H}_n), \; 
     C^* = \sum_{i=1}^n  w_i \bigg( \dot{C}_{\hat{\theta}_{\tau}} ^2  + (C_{\hat{\theta}_{\tau}} - \hat{H}_n) \ddot{C}_{\hat{\theta}_{\tau}}  \bigg),  \; 
     w_i = \frac{(n+1)^2(n+2)}{i(n-i+1)}.
\end{equation*}

\subsection{Additional simulation results}
\begin{figure}[t]
\centering
\includegraphics[scale=0.7]{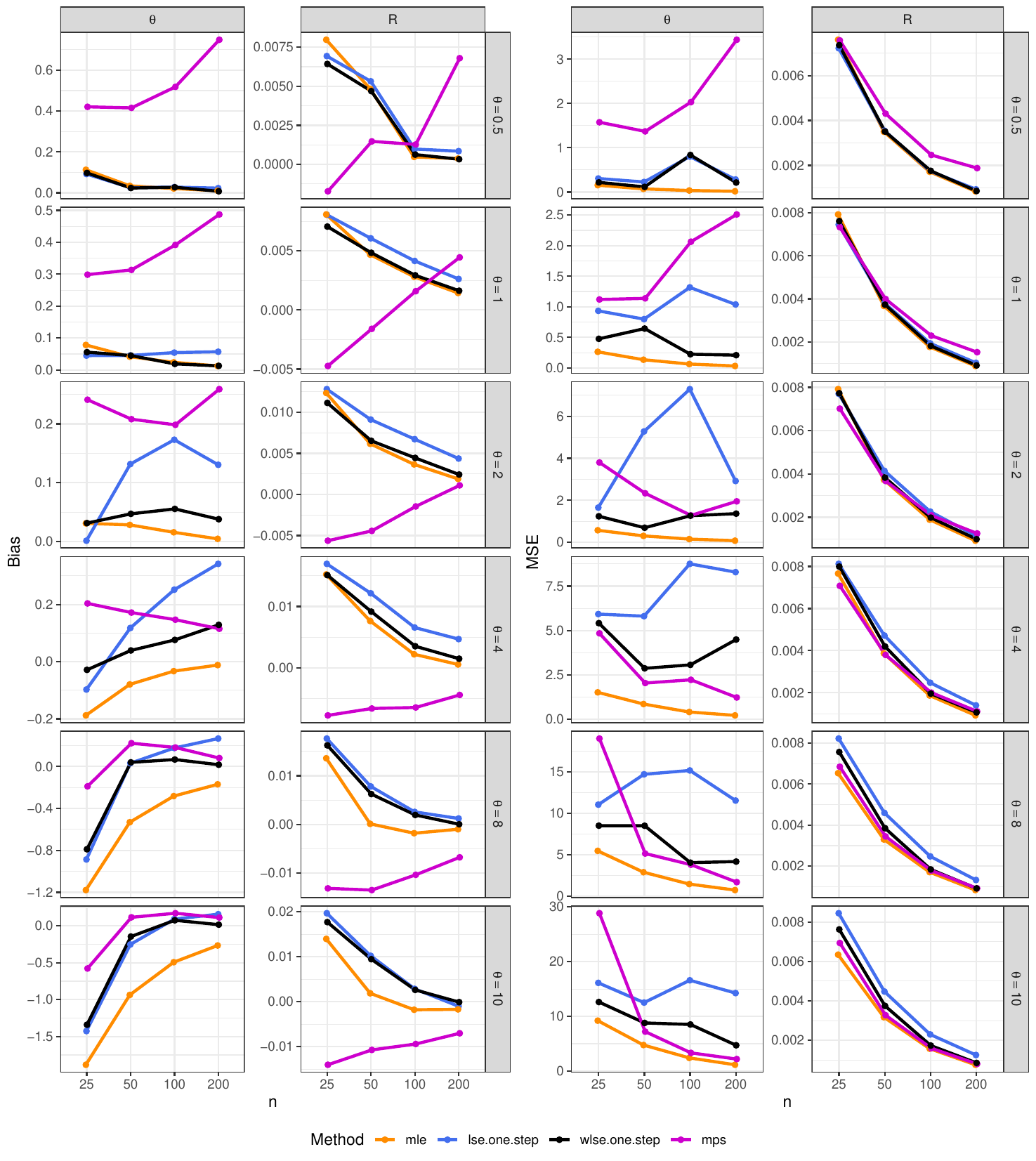}
\caption{Bias and MSE of the copula parameter $\theta$ and the stress-strength reliability $R$} 
\label{bias_mse_Rtheta_r1}
\end{figure}

\begin{figure}[t]
\centering
\includegraphics[scale=0.7]{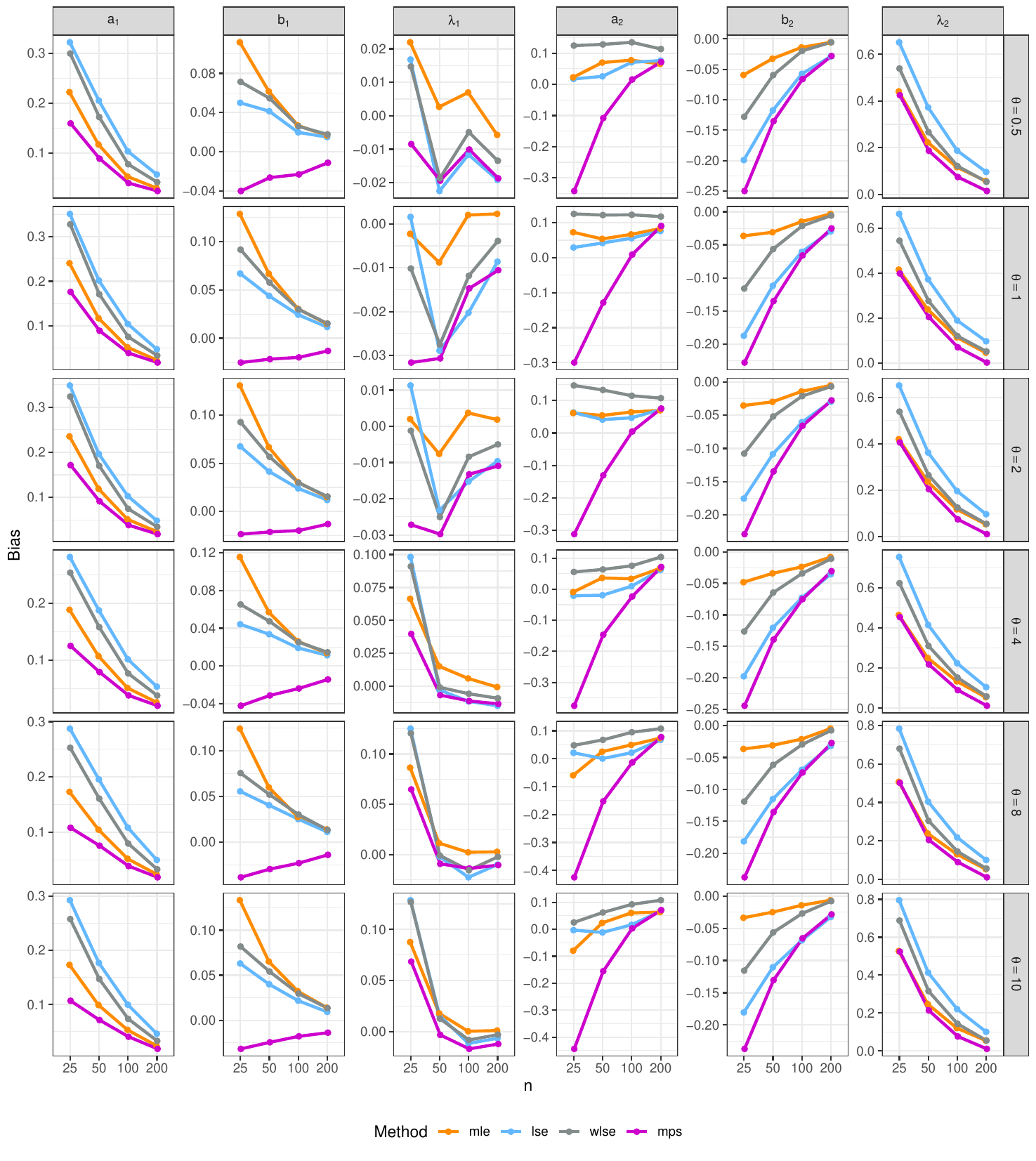}
\caption{Bias of the stress and strength distribution parameters} 
\label{bias_plot_parameters}
\end{figure}

\begin{figure}[t]
\centering
\includegraphics[scale=0.7]{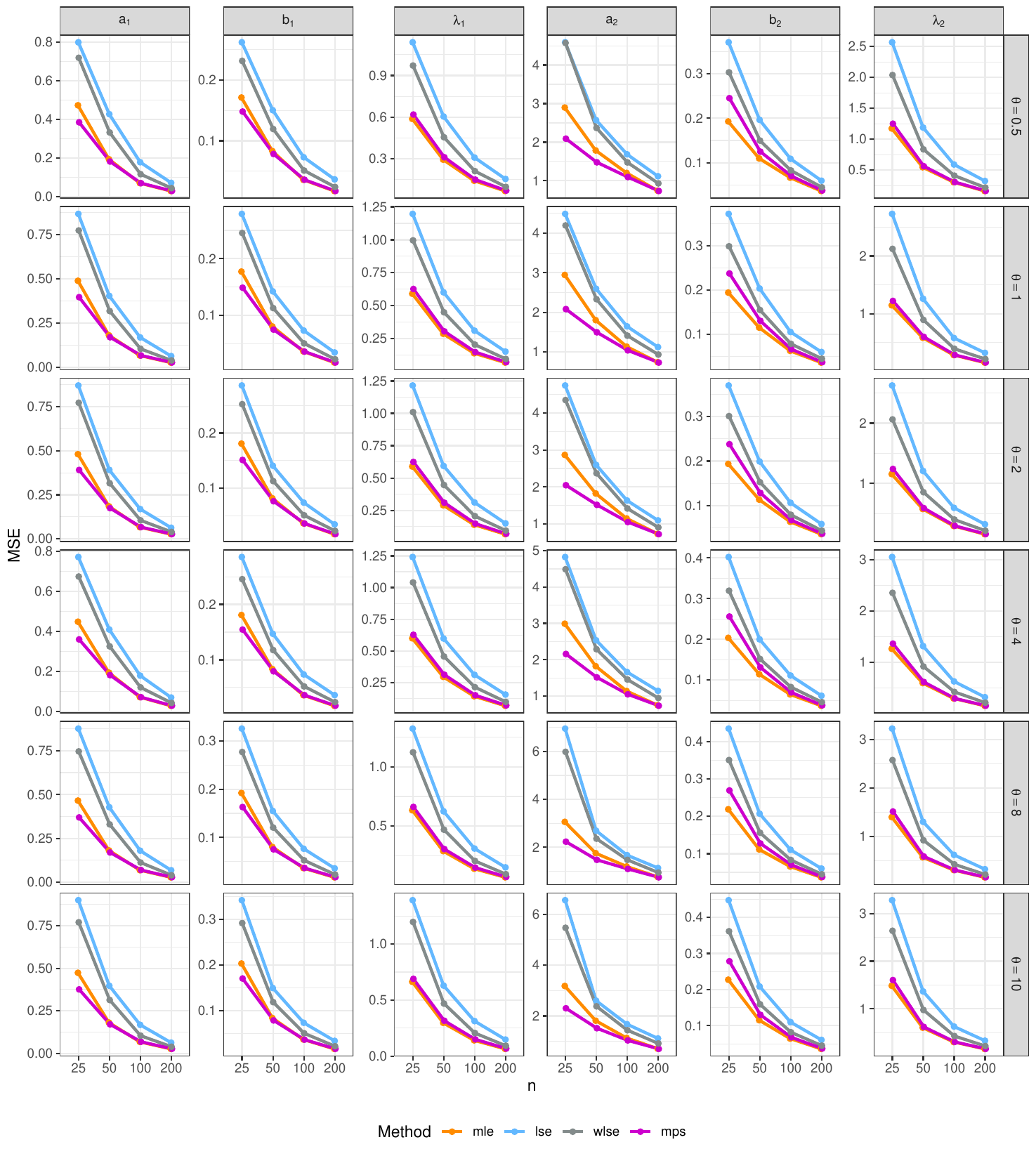}
\caption{MSE of the stress and strength distribution parameters}
\label{mse_plot_parameters}
\end{figure}

\clearpage

\bibliographystyle{unsrtnat}

\end{document}